\documentclass[a4paper,11pt]{article}

\usepackage{jcappub} % for details on the use of the package, please
\usepackage[export]{adjustbox} 
\usepackage[utf8]{inputenc}
\usepackage{color}
\usepackage{graphicx}
\usepackage{amsmath,amsfonts,amssymb}
\usepackage{acronym}
\usepackage{xspace}
\usepackage{multirow}
\usepackage{tabularx}
\usepackage{orcidlink}
\usepackage{ragged2e}
\usepackage{enumitem}
\usepackage{setspace}
\usepackage{caption}
\usepackage{subcaption}
\usepackage{booktabs}
\usepackage{comment}
\usepackage{cleveref}
\usepackage{listings}
\usepackage[ruled,vlined]{algorithm2e}

\SetCommentSty{mycommfont}

\newcommand{\bea}{\begin{equation}\begin{aligned}} 
\newcommand{\eea}{\end{aligned}\end{equation}}
\newcommand{\be}{\begin{equation}}
\newcommand{\ee}{\end{equation}}

\definecolor{rossocorsa}{rgb}{0.83, 0.0, 0.0}
\definecolor{tardisblue}{rgb}{0.0, 0.18, 0.53}
\definecolor{TardisBlueAcceso}{RGB}{0, 80, 200}

\title{Probing non-Gaussianity during reheating with SIGW in the LISA band}

\author[a]{Gabriele Perna\orcidlink{0000-0002-7364-1904}}
\author[b,c]{and Guillem~Domènech\orcidlink{0000-0003-2788-884X}}

\affiliation[a]{Keemilise ja Bioloogilise F\"u\"usika Instituut, R\"avala pst. 10, 10143 Tallinn, Estonia}
\affiliation[b]{Institute for Theoretical Physics, Leibniz University Hannover, Appelstraße 2, 30167 Hannover, Germany}
\affiliation[c]{Max Planck Institute for Gravitational Physics, Albert Einstein Institute, 30167 Hannover, Germany.}

\emailAdd{gabriele.perna@kbfi.ee}
\emailAdd{guillem.domenech@itp.uni-hannover.de}

\abstract{
We analyse the effects of a non-standard evolution of the Universe during the reheating epoch on the spectrum of scalar-induced gravitational waves (SIGWs), accounting for primordial non-Gaussianity. We show that given values of $w$ and $c_s^2$ leave characteristic features in the spectrum which can be detectable by third-generation interferometers like LISA. In addition, we argue that the specific reheating dynamics can suppress or even enhance the spectrum, with crucial consequences for its detectability. We perform a Fisher forecast for different values of $w$ and conduct scans to assess the signal's detectability for different amplitudes and central frequencies.
}

\begin{document}
\maketitle
\flushbottom

\section{Introduction}
Gravitational Waves (GWs) are a fundamental probe to shed light on the evolution of the Universe, whether they are produced during its late or early phases. A wide range of astrophysical and cosmological processes predicts the production of GWs, covering a broad frequency range, which spans tens of orders of magnitude from $10^{-18}$ Hz for Cosmic Microwave Background (CMB)-related experiments up to very high frequencies ($\sim \rm MHz$ or even higher), see, e.g., \cite{Caprini:2018mtu,Regimbau:2011rp,Guzzetti:2016mkm} for reviews.

Current GW searches are providing promising hints of a GW background. For instance, Pulsar Timing Array (PTA) collaborations reported evidence for a Hellings-Downs pattern in the angular correlations of pulsar time-of-arrival data, which is characteristic of GWs~\cite{NANOGrav:2023gor, NANOGrav:2023hde,EPTA:2023fyk, EPTA:2023sfo, EPTA:2023xxk,Reardon:2023gzh, Zic:2023gta, Reardon:2023zen,Xu:2023wog}. In the coming decades, third-generation ground-based interferometers such as Einstein Telescope (ET)~\cite{Maggiore:2019uih,ET:2025xjr} and Cosmic Explorer (CE)~\cite{Reitze:2019iox}, and the space-based LISA~\cite{LISA:2017pwj}, will exquisitely probe respectively the $[1-10^3]$ Hz frequency band and the mHz band, where many different cosmological GW signals are plausible~\cite{ET:2025xjr,Maggiore:2007ulw,Guzzetti:2016mkm,Cai:2017cbj,Caprini:2018mtu}.\\ 

Among possible cosmological GW sources, we focus on scalar-induced gravitational waves (SIGWs) \cite{Tomita:1967wkp, Matarrese:1992rp, Bruni:1996im, Matarrese:1997ay, Acquaviva:2002ud, Carbone:2004iv, Ananda:2006af, Baumann:2007zm, Kohri:2018awv, Domenech:2020kqm, Pi:2020otn, Domenech:2021ztg, Adshead:2021hnm, Perna:2024ehx, Iovino:2025xkq}. These signals inevitably arise at second order in perturbation theory. While scalar, vector, and tensor perturbations decouple at linear order, second-order tensor modes can be generated by a quadratic source of first-order scalars (see also~\cite{Bari:2023rcw, Picard:2023sbz} for tensor-induced GWs). As a result, the amplitude and spectral shape of SIGWs are directly related to the properties of the primordial scalar power spectrum and, consequently, to the mechanism of inflation responsible for those features (see e.g. the recent analysis by \cite{LISACosmologyWorkingGroup:2025vdz}). 

Various inflationary models predict an enhancement of the power spectrum at LISA scales, both from single-field scenarios with specific features in the potential~\cite{Ivanov:1994pa,Kinney:1997ne,Inoue:2001zt,Kinney:2005vj,Martin:2012pe,Motohashi:2017kbs,Ozsoy:2019lyy, Karam:2022nym,Balaji:2022dbi,Allegrini:2024ooy,Allegrini:2025jha} as well as multi-field models~\cite{Garcia-Bellido:1996mdl, Kawasaki:2015ppx, Pi:2017gih, Kallosh:2022vha, Braglia:2022phb, Tada:2023pue,Tada:2023fvd,Wang:2024vfv,Iacconi:2024hmg}. Any enhancement in the scalar spectrum leads to a large induced GW signal, potentially detectable by present and future interferometers. In addition, mechanisms such as preheating or early matter-dominated eras can produce a strong enhancement of the GW spectrum~\cite{Jedamzik:2010hq,Domenech:2024wao,Pearce:2023kxp}. Therefore, SIGWs represent a powerful probe both for discriminating among early-Universe models and for reconstructing the small-scale scalar power spectrum generating them. 

A further important feature of SIGWs is their sensitivity to the expansion history of the Universe. Namely, the SIGW spectrum is computed by accounting for the evolution of tensor modes during the epoch considered and, therefore, acquires specific characteristic features depending on the background equation of state during that epoch, say $w$. In the standard Hot Big Bang cosmology, a radiation-dominated epoch ($w=1/3$) precedes a matter-dominated epoch ($w=0$), followed by a dark energy epoch (in which we are transitioning nowadays). However, the transition from an inflating Universe to the standard Hot Big Bang remains unexplored. Such a transition stage, often called reheating, could have different values of $w$, depending on the scenario considered (see, e.g., \cite{Allahverdi:2010xz} for a review). The GW frequencies probed by ET, CE and LISA precisely correspond to yet unexplored epochs in the very early Universe, which could well cover the transition to the standard Hot Big Bang cosmology. With SIGWs, we can test the particle content and equation of state of that phase (see also \cite{Yu:2024xmz,Domenech:2025bvr,Yu:2025cqu} for the effects of dissipation).

In this work, we take an agnostic approach and span values of $w$ in the range $[0,1]$, which includes scenarios such as cannibal dark matter models~\cite{Carlson:1992fn,Pappadopulo:2016pkp,Farina:2016llk,Erickcek:2020wzd} ($w\simeq0.1$), oscillating scalar fields in power-law potentials~\cite{Bernal:2019mhf,Garcia:2020eof, Garcia:2023dyf} ($w=0,1/3,1/2,\dots$), and chaotic inflation scenarios~\cite{Linde:1990flp,Felder:2000hr,Podolsky:2005bw}, where $w\in[0.2,0.3]$. The reheating temperature (defined as the temperature at the start of the Hot Big Bang cosmology) and its corresponding frequency scale play a crucial role in determining the amplitude of the SIGW spectrum relative to the standard radiation-dominated case. It can lead to either an enhancement or a suppression of the SIGW amplitude, depending on whether $w>1/3$ or $w<1/3$. On top of that, the value of the scalar sound speed has a non-negligible impact on the resulting SIGW spectrum~\cite{Domenech:2021ztg}.

Lastly, SIGWs are highly sensitive to primordial non-Gaussianity (NG), as they are directly sourced by higher-order correlators of the scalar perturbations \cite{Cai:2018dig,Unal:2018yaa,Yuan:2020iwf,Atal:2021jyo,Adshead:2021hnm,Li:2023qua,Li:2023xtl,Perna:2024ehx,Li:2025met,Zeng:2025cer}. Primordial NG represents a powerful probe of the early Universe, being sensitive to the underlying dynamics and interactions. Different inflationary scenarios predict different levels of NG~\cite{Komatsu:2010hc,Babich:2004gb,Bartolo:2004if}. Current observations have placed tight constraints on the NG parameter $f_{\rm NL}$ on large scales \cite{Planck:2019kim}, but the small scales remain essentially unconstrained and could be probed by SIGWs~\cite{Iovino:2025xkq}. Different works have analysed the effects of NG on the SIGW spectrum in the radiation-dominated universe~\cite{Unal:2018yaa, Adshead:2021hnm, Perna:2024ehx} and limits in current analysis when NG is considered~\cite{Iovino:2024sgs}. Here, we generalise previous work by including an arbitrary, constant equation of state $w$. We forecast LISA's capability to detect the SIGW spectrum produced during a reheating epoch with a constant equation of state, including local NG.

The paper is organised as follows.  In \cref{sec:Derivation}, we derive the GW spectrum for a general $w$, and we discuss the role of primordial NG.  In \cref{sec:results}, we briefly discuss some benchmark reheating models, present the spectra of SIGWs for different $w$ and $c_s^2$, and the results of a Fisher forecast with some scans in the parameter space. We conclude in \cref{sec:conclusion}. Details of the calculations are in the appendices.

%%%%%%%%%%%%%%%%%%%%%%%%%%%%%%%%%%%%%%%%%%%
\section{Derivation of the spectra}
\label{sec:Derivation}
Our main goal is to assess LISA's capability to observe SIGWs produced during a non-standard evolutionary history. High detectability and meaningful parameter estimation depend on the amplitude of the SIGW spectrum and the specific imprint each parameter leaves on it. To obtain these dependencies, we present below a self-consistent review of the derivation, including NG and a general equation of state.

First, we need to solve the equation of motion for tensor fluctuations at second order. Following~\cite{Domenech:2021ztg,Adshead:2021hnm,Perna:2024ehx}, our starting point is a perturbed Friedmann–Lemaître–Robertson–Walker (FLRW) metric in the Newton gauge given by
\bea
\label{eq:metric}
    ds^2 = \hspace{0.2cm} a^2(\eta)\Bigg\{&-\left(1+2\Phi^{(1)}+\Phi^{(2)}\right)d\eta^2\\
    &+ 2\omega_i^{(2)}d\eta dx^i + \Big[(1-2\Psi^{(1)}-\Psi^{(2)})\delta_{ij}
    + \frac{1}{2}h_{ij}\Big]dx^idx^j\Bigg\}\,,
\eea
where we consider only second order tensors, that is $h_{ij}=h_{ij}^{(2)}$. To keep the calculation as general as possible, we will leave the equation of state $w$ and the speed of sound of scalar fluctuations $c_s$  as free parameters, only assuming that in the period of interest, they are (sufficiently) constant. After some calculation, the Einstein-Equations at second order for tensor modes read~\cite{Acquaviva:2002ud, Mollerach:2003nq, Ananda:2006af, Baumann:2007zm}
\begin{equation}
h_{ij}''(\eta,\vec{ x})
+2\mathcal H h_{ij}'(\eta,\vec{ x})
-\nabla^2 h_{ij}(\eta,\vec{ x})
=-4 \mathcal T_{ij}{}^{lm} S_{lm}(\eta,\vec{ x}),
\label{eq:eom GW}
\end{equation}
where a prime denotes derivative with respect to conformal time $\eta$, $\mathcal H=a'/a$ denotes the conformal Hubble parameter, $\mathcal{T}_{ij}^{lm}$ is a projector in the TT gauge, and $ S_{ij}$ is the source term given by
\begin{equation}
 S_{ij} = 4\Phi\partial_i\partial_j\Phi+2\partial_i\Phi\partial_j\Phi-\frac{4}{3(1+w)}\partial_i\left(\frac{\Phi'}{\mathcal H}+\Phi\right)\partial_j\left(\frac{\Phi'}{\mathcal H}+\Phi\right)\,.
\label{eq:source S}
\end{equation}

A common approach to solving Eq.~\eqref{eq:eom GW} is to work in Fourier space, obtaining
\be
\label{eq:EOM_tensors}
h_{k,\lambda}''+2\mathcal{H}h'_{k,\lambda}+k^2h_{k,\lambda}=\mathcal{S}_{k,\lambda}\,,
\ee
where $\lambda$ denotes the GW polarization and $\mathcal{S}_{k,\lambda}$ is the Fourier transform of the source function. Its expression can be obtained considering that $e_{ij}^\lambda\mathcal{T}_{lm}{}^{ij}=e_{lm}^\lambda$, where $e_{ij}^\lambda(\hat{k})$ are the polarization tensors of a GW propagating along the $\hat {k}$ direction and which satisfy $e^\lambda_{ij}(\hat{k})k^i=0$. We take the normalization that  $e^\lambda_{ij}(\hat{k})(e^{\lambda'ij}(\hat{k}))^*=\delta^{\lambda\lambda'}$. With this, we arrive at
\be
\mathcal{S}_{k,\lambda}=4\int \frac{d^3q}{(2\pi)^3} e_{\lambda}^{ij} q_i q_j \Phi(\vec{q}) \Phi(\vec{k}-\vec{q})f({q},|\vec{k}-\vec{q}|,\eta)\,,
\ee
where
\bea
\label{eq:func_kqeta}
f(k,q,\eta) = &\bigg[2 T_\Phi(q,\eta)T_\Phi(|k-q|,\eta).\\
&\left.+\frac{4}{3(1+w)} \left(\frac{T_\Phi'(q,\eta)}{\mathcal{H}}+T_\Phi(q,\eta)\right)\left(\frac{T_\Phi'(|\vec{k}-\vec{q}|,\eta)}{\mathcal{H}}+T_\Phi(|k-q|,\eta)\right)\right]\,.
\eea
In the last equation we introduced the transfer function for $\Phi$, $T_\Phi$, whose explicit expression for generic $w$ and $c_s^2$ is reported in app. \ref{app:scalars_eom}. As argued in app. \ref{app:Greens_tensors}, \cref{eq:EOM_tensors} can be solved using the Green's function method which leads to
\bea\label{eq:hsolintroI}
h_{\lambda}(k,\eta) = & \int_0^\eta d\bar{\eta}G_{k,\lambda}(\eta,\bar{\eta}) S_\lambda(\bar{\eta},k)\nonumber\\
= & 4 \int_0^\eta d\bar{\eta}G_{k,\lambda}(\eta,\bar{\eta}) \int \frac{d^3q}{(2\pi)^3} e_{\lambda}^{ij} q_i q_j \Phi(\vec{q}) \Phi(\vec{k}-\vec{q})f({q},|\vec{k}-\vec{q}|,\eta)\\
\equiv & 4\int\frac{d^3q}{(2\pi)^3} e_{\lambda}^{ij} q_i q_j \Phi(\vec{q}) \Phi(\vec{k}-\vec{q}) I(q,|\vec{k}-\vec{q}|,\eta)\,.
\eea
In the last line, we introduced the kernel function $I(q,|\vec{k}-\vec{q}|,\eta)$ (defined later in eq.~\eqref{eq:kerneldef}), which accounts for the evolution of scalar perturbations during the specific epoch of interest. 

Now, it is important to specify that different, and somewhat similar, definitions of the kernel are adopted in the literature, as we explain in more detail below. In this work, we define it when $h_\lambda$ is still expressed in terms of $\Phi$ for simplicity. One can then use that, on superhorizon scales, $\Phi$ is related to the curvature perturbation on uniform density slices $\zeta$ via
\be
\Phi(\vec{k})=\frac{3(1+w)}{5+3w}\zeta(\vec{k})=\frac{(2+b)}{(3+2b)}\zeta(\vec{k})\,,
\ee
where, for later convenience, we introduced
\be
b=\frac{1-3w}{1+3w} \quad\Leftrightarrow\quad w=\frac{1-b}{3(1+b)}\,.
\ee
One can then define a new kernel using $\zeta$ as $\mathbb{I}(q,|\vec{k}-\vec{q}|,\eta)=\frac{(2+b)^2}{(3+2b)^2}I(q,|\vec{k}-\vec{q}|,\eta)$, which we will use later on for eventual comparisons with other works (see app.~\ref{app:comparison}). 

With the Green's function solution and the kernels, we compute the strain two-point correlator as
\bea
\label{eq:tensor_h_sigw}
\langle h_{\lambda}(\eta,\vec{k})h_{\lambda'}(\eta,\vec{k}')\rangle = & 2^4 \int\frac{d^3q}{(2\pi)^3}\frac{d^3q'}{(2\pi)^3} Q_{\lambda}(\vec{k},\vec{q})Q_{\lambda'}(\vec{k}',\vec{q}') \\
& \times I(q,|\vec{k}-\vec{q}|,\eta)I(q',|\vec{k'}-\vec{q'}|,\eta) \langle\Phi(\vec{q}) \Phi(\vec{k}-\vec{q}) \Phi(\vec{q'}) \Phi(\vec{k'}-\vec{q'}) \rangle\,,
\eea
or equivalently,
\bea
\label{eq:tensor_h_sigw_2}
\langle h_{\lambda}(\eta,\vec{k})h_{\lambda'}(\eta,\vec{k}')\rangle = & 2^4 \int\frac{d^3q}{(2\pi)^3}\frac{d^3q'}{(2\pi)^3} Q_{\lambda}(\vec{k},\vec{q})Q_{\lambda'}(\vec{k}',\vec{q}') \\
& \times \mathbb{I}(q,|\vec{k}-\vec{q}|,\eta)\mathbb{I}(q',|\vec{k'}-\vec{q'}|,\eta) \langle\zeta(\vec{q}) \zeta(\vec{k}-\vec{q}) \zeta(\vec{q'}) \zeta(\vec{k'}-\vec{q'}) \rangle\,,
\eea
with $Q_{\lambda}(\vec{k},\vec{q})=e_\lambda^{ij}q_iq_j$. Note that $Q_{\lambda}(\vec{k},\vec{q})$ is called the projection factor and will be briefly analysed below (its properties can be found, e.g. in~\cite{Adshead:2021hnm, Perna:2024ehx}). 

With the definition of tensor power spectrum $P_{h,\lambda}$, namely
\be
\label{eq:tensor_ps}
\langle h_{\lambda}(\eta,\vec{k})h_{\lambda'}(\eta,\vec{k}')\rangle = (2\pi)^3\delta(\vec{k}+\vec{k}')P_{h,\lambda}(k,\eta)\delta_{\lambda,\lambda'}\,,
\ee
we may write the GW energy density as
\bea
    \label{eq:omega}
    \Omega_{\rm GW}(k,\eta) \equiv \frac{\rho_{\rm GW}(k,\eta)}{\rho_{tot}(\eta)} 
    {=} & \frac{1}{48} x^2(1+b)^{-2}\frac{k^3}{2\pi^2}\sum_{\lambda=+,\times} \overline{P_{h,\lambda}(k,\eta)}\,,
\eea
where we defined $x\equiv k\eta$.
Lastly, we map the GW energy density at the emission to the one observed today at frequency $f$ by
\be
    \Omega_{\rm GW,0}(f) =  \Omega_{\rm rad,0} \left(\frac{g_*(T)}{g_*(T_0)}\right)  \left(\frac{g_S(T)}{g_S(T_0)}\right)^{-\frac{4}{3}} \Omega_{\rm GW}(f,\eta)\,,
\ee
where $g_*(T)$ and $g_S(T)$ are the energy and entropy degrees of freedom at temperature $T$ (note that the factor involving those quantities is sometimes called $c_g(T)$~\cite{Espinosa:2018eve}), while $\Omega_{\rm rad,0}$ is the radiation energy density today.

In this work, we consider SIGW produced during a reheating stage before the usual radiation domination epoch and therefore need to account for the effects of the transition between the two stages on the GW spectrum. For analytical simplicity, we will consider an instantaneous transition. In this case, as shown by~\cite{Domenech:2021ztg}, this can be done by matching the kernels at the time of the transition between the radiation and reheating stages at $\eta=\eta_{\rm RH}$ (see also \cite{Altavista:2023zhw}). This procedure exploits the continuity of $h_{ij}$ and of its derivative (and as a consequence the continuity of the kernel $I$), but neglects any additional sourcing of SIGWs after the transition. This is a good approximation as long as we consider SIGWs that enter the horizon well inside the reheating phase \cite{Domenech:2021ztg}. Thus, we always consider $f>f_{\rm RH}$.

Before entering into details, let us explain the qualitative differences of a general $w$ stage relative to the SIGW spectrum in radiation domination.  The SIGW spectrum produced during the reheating stage then evolves during the standard radiation-dominated epoch.  For modes entering well before the reheating scale, the final SIGW spectrum can be obtained by computing it as we explain in \cref{sec:spectra_eval} and evaluating it at the end of reheating $\eta_{\rm RH}$. To do a quick estimate, we note that the factor $x^{-2b}$, which appears in the final expression for $\Omega_{\rm GW}$ (see Eq.~\eqref{Eq::Omega_G}), evaluated at the end of reheating, reads
\be
\label{eq:frh_dep}
    x^{-2b}_{\rm RH} = \left(\frac{k}{k_{\rm RH}}\right)^{-2b}(1+b)^{-2b} = \left(\frac{k_*}{k_{\rm RH}}\right)^{-2b}\left(\frac{k}{k_{*}}\right)^{-2b}(1+b)^{-2b}\,,
\ee
we introduced $k_*$ as a reference comoving scale. For a peaked primordial spectrum, we take $k_*$ to be the peak scale.  Note that, for consistency, we must have that $k_*/k_{\rm RH}> 1$.
Eq.~\eqref{eq:frh_dep} allows us to write the SIGW spectrum in terms of the dimensionless quantity $\kappa=k/k_*$, at the price of introducing an additional multiplicative factor which goes like the ratio $k_*/k_{\rm RH}$. All of the spectra shown in \cref{sec:results} below are normalised over this ratio, in order to be independent of the possible choice of the reheating scale. For later use, we report here the relation between the reheating frequency $f_{\rm RH}=k_{\rm RH}/2\pi$ and the reheating temperature $T_{\rm RH}$, that is \cite{Domenech:2021ztg}
\begin{align}\label{eq:krh}
f_{\rm RH}=1.3\times10^{-5}\,{\rm Hz}\,\left(\frac{T_{\rm RH}}{10^3\,{\rm GeV}}\right)\left(\frac{g_*(T_{\rm RH})}{106.75}\right)^{1/2}\left(\frac{g_{S}(T_{\rm RH})}{106.75}\right)^{-1/3}\,.
\end{align}
The higher the reheating temperature, the higher the reheating scale.  Note that the reheating temperature is bounded from below by Big Bang Nucleosynthesis (BBN) constraints, yielding $T\gtrsim4 \,{\rm MeV}$ \cite{Kawasaki:1999na,Kawasaki:2000en,Hannestad:2004px,Hasegawa:2019jsa}. Thus, we consider $f_{\rm RH}>10^{-10}~\mathrm{Hz}$.

From Eq.~\eqref{eq:frh_dep}, we also see the effects of the reheating scale on the spectrum. First, recall that $b=0$ for $w=1/3$, $b<0$ for $w>1/3$ and $b>0$ for $w<1/3$. Then, we see that given a fixed value of $k_*/k_{\rm RH}$, the resulting spectrum is boosted for $b<0$ and a lower amplitude of the primordial spectrum is needed to generate a detectable GW signal. On the other hand, for $b>0$, the signal is suppressed, and a higher amplitude is needed to detect it. This effect can be understood from a physical point of view by considering that the energy density of GWs evolves as $\rho_{\rm GW}\propto a^{-4}$ while the background goes as $\rho_{\rm bg}\propto a^{-3(1+w)}$, and thus it follows that during the reheating phase $\Omega_{\rm GW}\propto a^{3w-1}$. At horizon crossing, that is when $a_kH_k=k$ for a given comoving wavenumber $k$, we find that $k\propto a_k^{-(1+3w)/2}$. Therefore, using the reference scale $k_*$, we can write
\be
\Omega_{\rm GW}^{\rm RH} = \Omega_{\rm GW}^{*} \left(\frac{a_{\rm RH}}{a_{*}}\right)^{3w-1} = \Omega_{*} \left(\frac{k_{*}}{k_{\rm RH}}\right)^{-2\frac{1-3w}{1+3w}} = \Omega_{*} \left(\frac{k_{*}}{k_{\rm RH}}\right)^{-2b}\,,
\ee
which agrees with the result we find later. Note that the enhancement for $b<0$ is relative, as the energy density of GWs is diluted regardless of the background dynamics.

%%%%%%%%%%%%%%%%%%%%%%%
\subsection{Primordial non-Gaussianity}
Primordial NG characterises the statistical properties of primordial perturbations and is crucial for shedding light on the physics of the very early Universe (see~\cite{Komatsu:2010hc,Babich:2004gb,Bartolo:2004if}).  As shown in \cref{eq:tensor_h_sigw,eq:tensor_h_sigw_2}, the spectrum of SIGWs is sensitive to the presence of NG. Namely, the induced tensor spectrum depends on the trispectrum of primordial fluctuations (the Fourier transform of the four-point correlation function). 

In general, the trispectrum can be written as the sum of two contributions, the so-called connected and disconnected trispectrum, as
\be
\label{eq:4pf}
\mathcal{T}_\zeta( \vec{k}_1,\vec{k}_2,{\vec{k}_3},{\vec{k}_4})=\mathcal{T}_\zeta( \vec{k}_1,\vec{k}_2,{\vec{k}_3},{\vec{k}_4})|_c+\mathcal{T}_\zeta( \vec{k}_1,\vec{k}_2,{\vec{k}_3},{\vec{k}_4})|_d\,.
\ee
For a Gaussian field, the connected contribution vanishes, and the disconnected one can be written as a product of two power spectra. In the presence of NG, the connected piece may dominate in the ultra-violate (UV) regime of the GW spectrum. This decomposition makes it clear that the presence of NG leaves specific imprints on the SIGW spectrum and may allow current and future GW interferometers to detect or constrain primordial NG.

A common, and simple, way to model primordial NG is to adopt a polynomial expansion, where the non-Gaussian field $\zeta$ is written in terms of a Gaussian field $\zeta_{\rm G}$ as
\be
\label{eq:local_exp}
\zeta(\vec{x}) = \zeta_{\rm G}(\vec{x}) + \frac{3}{5} f_{\rm NL} (\zeta_{\rm G}^2(\vec{x})-\langle \zeta_{\rm G}^2\rangle)\,,
\ee
with $f_{\rm NL}$ quantifying the amount of primordial NG. Note that such an expansion is originally inspired to parametrise perturbative departures from a Gaussian field and, although it may be taken at face value, one must be aware of its limitations. For instance, as argued by~\cite{Iovino:2024sgs} it may miss the full effect due to non-linearities in the evaluation of the SIGW spectrum and, as analysed by~\cite{Garcia-Saenz:2022tzu,Iovino:2025cdy}, the perturbativity condition is crucial for this relation to hold, and its range of validity is strongly model-dependent (we refer to also~\cite{Veermae:2026yzz} for a full non-perturbative approach and to ~\cite{Caravano:2026hca} for a recent complementary approach based on lattice simulations). Nevertheless, as a first study of the impact of NG in the SIGW spectrum for a general equation of state, we focus on local NG alone, as this allows us to qualitatively understand its imprints and detectability. We leave a more general study for future work.

%%%%%%%%%%%%%%%%%%%%
\subsection{Evaluation of the kernel}
The explicit calculation of the kernel, introduced in eq.~\eqref{eq:hsolintroI}, is a necessary step to obtain the SIGW power spectrum for the non-Gaussian contribution. To do so, we first start from the definition of the kernel, namely
\bea\label{eq:kerneldef}
I(v,u,x) = & \int_0^\eta d\bar{\eta}\, G(\eta,\bar{\eta})f(v,u,\bar{x})\,,
\eea
where the Green's function, $G(\eta,\bar{\eta})$, is given in \cref{eq:green_k,eq:func_kqeta}, and $f(v,u,\bar{x})$ in \cref{eq:func_kqeta}. Now, using the expressions of the scalar transfer functions reported in app.~\ref{app:scalars_eom} and introducing $v=|\vec{q}|/k$ and $u=|\vec{k}-\vec{q}|/k$, we obtain\footnote{We add that the definition of the kernel changes throughout the literature. It is frequently written without the factor $1/k^2$, since it simplifies with a $k^2$ coming from the explicit expression of $Q_\lambda(\vec{k},\vec{q})$.}~\cite{Domenech:2019quo,Domenech:2020kqm}
\bea\label{eq:kerneldef}
I(v,u,x) =  & \int_0^\eta d\eta \frac{\pi}{k} \frac{(k\bar{\eta})^{b+\frac{3}{2}}}{(k{\eta})^{b+\frac{1}{2}}} \frac{2^{2b+3}\Gamma^2\left[b+\frac{5}{2}\right]}{(2+b)(3+2b)} (c_s^2 u v)^{-b-\frac{1}{2}}\bar{x}^{-2b-1}\\
& \times \left(J_{b+\frac{1}{2}}(k\bar{\eta})Y_{b+\frac{1}{2}}(k{\eta})-Y_{b+\frac{1}{2}}(k\bar{\eta})J_{b+\frac{1}{2}}(k{\eta})\right)\\
& \times \left[J_{b+\frac{1}{2}}(c_s v \bar{x})J_{b+\frac{1}{2}}(c_s u \bar{x})+\frac{b+2}{b+1}J_{b+\frac{5}{2}}(c_s v \bar{x})J_{b+\frac{5}{2}}(c_s u \bar{x})\right]\\
= & \frac{1}{(b+2)(3+2b)}\frac{2^3}{k^2} \pi \Gamma^2\left[b+\frac{5}{2}\right]4^b(c_s^2 u v x)^{-b-\frac{1}{2}}\left[Y_{b+\frac{1}{2}}(x)\mathcal{I}_Y-J_{b+\frac{1}{2}}(x)\mathcal{I}_J\right]\,.
\eea
Note that, for compactness, we introduced in the last step the integrals 
\bea
\mathcal{I}_{J/Y} = \int_0^x d\bar{x} \,\bar{x}^{\frac 1 2 -b} 
\begin{Bmatrix}
J_{b+\frac{1}{2}}(\bar{x})\\
Y_{b+\frac{1}{2}}(\bar{x})
\end{Bmatrix}
\left(J_{b+\frac{1}{2}}(c_s v \bar{x})J_{b+\frac{1}{2}}(c_s u \bar{x})+\frac{b+2}{b+1}J_{b+\frac{5}{2}}(c_s v \bar{x})J_{b+\frac{5}{2}}(c_s u \bar{x})\right)\,.
\eea

Fortunately, the integrals $\mathcal{I}_{J/Y}$ can be solved analytically~\cite{Domenech:2019quo,Domenech:2020kqm,Domenech:2021ztg} for a constant $b$ and $c_s$ leading to
\bea
\label{eq:general_kernel}
  I(v,u,x) &= \frac{2}{k^2}\frac{(3+2b)}{(b+2)} \Gamma^2\left[b+\frac{3}{2}\right]4^b x^{-b-1}\frac{|1-y^2|^{\frac{b}{2}}}{c_s^2 u v}\\
    &\times\left\{ \Theta[c_s(u+v)-1]\left[ -\cos\left(x-\frac{b\pi}{2}\right)  \left(\mathsf{P}_b^{-b}(y)+\frac{b+2}{b+1}\mathsf{P}_{b+2}^{-b}(y)\right)\right.\right.\\
    &\left.\left.\qquad\qquad\qquad\qquad\qquad+\sin\left(x-\frac{b\pi}{2}\right) \frac{2}{\pi}\left(\mathsf{Q}_b^{-b}(y)+\frac{b+2}{b+1}\mathsf{Q}_{b+2}^{-b}(y)\right) \right]\right.\\
    &\left.\quad- \Theta[1-c_s(u+v)]\frac{2}{\pi}\left[\sin\left(x-\frac{b\pi}{2}\right)\left(\mathcal{Q}_{b}^{-b}(-y)+2\frac{b+2}{b+1}\mathcal{Q}_{b+2}^{-b}(-y)\right)\right]\right\}\,,
\eea
with
\be
y = 1-\frac{1-c_s^2(u-v)^2}{2c_s^2 u v}\,.
\ee
In \cref{eq:general_kernel}, $\mathsf{P}^{\mu}_{\nu}$ and  $\mathsf{Q}^{\mu}_{\nu}$ are the Ferrer's functions and ${\cal Q}^{\mu}_{\nu}$ the Olver's function (their expression for specific values of $b$ can be found in app.~\ref{app:Olv_Ferr}).

We are now ready to compute the oscillation-averaged kernel that enters the SIGW spectrum in the presence of general NG, \cref{eq:tensor_h_sigw}. This is given by
\bea
\label{eq:generalized_kernel}
& \overline{I(v_1,u_1,x\gg1)I(v_2,u_2,x\gg1)} =\frac{2}{k^4}\frac{(3+2b)^2}{(b+2)^2} \Gamma^4\left[b+\frac{3}{2}\right]\frac{4^{2b} x^{-2b-2}}{c_s^4}\frac{|1-y^2_1|^{{b/2}}}{ u_1v_1 }\frac{|1-y^2_2|^{{b/2}}}{u_2v_2}\\
&\times\left\{ I_A(y_1,u_1,v_1,c_s,b)I_A(y_2,u_2,v_2,c_s,b)+ \frac{4}{\pi^2} I_B(y_1,u_1,v_1,c_s,b)I_B(y_2,u_2,v_2,c_s,b)\right\}\,,
\eea
with
\bea
    I_A(y_i,u_i,v_i,c_s,b) = & \left(\mathsf{P}^{-b}_{b}(y_i)+\frac{b+2}{b+1}\mathsf{P}^{-b}_{b+2}(y_i)\right)\Theta[c_s(u_i+v_i)-1]\\
    I_B(y_i,u_i,v_i,c_s,b) = & \left(\mathsf{Q}^{-b}_{b}(y_i)+\frac{b+2}{b+1}\mathsf{Q}^{-b}_{b+2}(y_i)\right)\Theta[c_s(u_i+v_i)-1]\nonumber\\
    &-\left({\cal Q}^{-b}_{b}(-y_i)+2\frac{b+2}{b+1}\mathsf{\cal Q}^{-b}_{b+2}(-y_i)\right)\Theta[1-c_s(u_i+v_i)]\,.
\eea
The overbar indicates an oscillation average which allows us to replace $\sin^2x=\cos^2x\to 1/2$ and $\sin x \cos x \to 0$.
We recover the Gaussian case \cite{Domenech:2021ztg} by setting $v_1=v_2$ and $u_1=u_2$ in eq.~\eqref{eq:generalized_kernel}, namely
\bea
\overline{I(v_1,u_1,x\gg1)^2} = & \frac{2}{k^4}\frac{(3+2b)^2}{(b+2)^2} \Gamma^4\left[b+\frac{3}{2}\right]4^{2b} x^{-2b-2}\frac{|1-y^2|^{{b}}}{c_s^4 u^2 v^2}\\
&\times\left\{ I_A^2(y_1,u_1,v_1,c_s,b)+ \frac{4}{\pi^2} I_B^2(y_1,u_1,v_1,c_s,b)\right\}\,.
\eea
Our results agree with the analytical kernels computed in radiation domination by~\cite{Kohri:2018awv} in the Gaussian case and by~\cite{Adshead:2021hnm,Perna:2024ehx} in the non-Gaussian one (see also~\cite{Domenech:2021ztg,LISACosmologyWorkingGroup:2025vdz}).

\subsection{Projection Factor}
We are left with the evaluation of the projection factor defined as
\be
Q_{\lambda}(\vec{k},\vec{q})=e_\lambda^{ij}q_iq_j\,,
\ee
see also eq.~\eqref{eq:tensor_h_sigw} and below.
We obtain its explicit expression without loss of generality by orienting $\vec{k}$ along the $\hat{z}$ axis, and so we may write $\vec{q_i}$ as
\bea
    \vec{q_i} &= q_i(\cos\phi_i \sin\theta_i,  \sin\phi_i \sin\theta_i, \cos\theta_i)\,.
\eea
For the plus and cross polarisation, we have that
\be
    Q_{\lambda}(\vec{k},\vec{q}) = \frac{q^2}{\sqrt{2}}\sin^2\theta\times
    \begin{cases}
    \cos2\phi, \hspace{0.2cm} \lambda = + \\
    \sin2\phi, \hspace{0.2cm}\lambda = \times
    \end{cases}\,.
\ee
Summing over the polarisations, one obtains
\bea
    \sum_{\lambda=+,\times}\Big[Q^2_{\lambda}(\vec{k},\vec{q})\Big] = \frac{q^4}{2}\sin^4\theta
\eea
and
\be
    \sum_{\lambda=+,\times}\left[Q_{\lambda}(\vec{k},\vec{q}_1)Q_{\lambda}(\vec{k},\vec{q}_2)\right]
     = \hspace{0.1cm} \frac{q_1^2q_2^2}{2}\sin^2\theta_1\sin^2\theta_2\cos(2(\phi_1-\phi_2))\,.
\ee
Further details can be found in~\cite{Adshead:2021hnm,Perna:2024ehx}.
%%%%%%%%%%%%%%%%%%%%%%%%%%%

%%%%%%%%%%%%%%%%%%%%%%%%%%%%%%%%%%%%%%%%%%%%%%%%%%%%%%%%%%%
\section{Results}
\label{sec:results}

In this section, we present our results for the Gaussian and local NG cases, and present a Fisher forecast for LISA. Although our formalism is valid for general values of $w$ and $c_s$ (with $c_s>0$), we select, when needed, a few values motivated by the existing literature, which we review below.

\subsection{Motivated reheating models}
\label{sec:reheating_models}
Reheating is defined as an intermediate stage between the end of inflation and the standard radiation-dominated era at the start of the hot Big Bang cosmology. Such a reheating phase remains so far unexplored; see, e.g., \cite{Allahverdi:2010xz} for a review of various possible reheating models. It is a usual approximation to assume that this transition is instantaneous, often called ``instantaneous reheating''. While this approximation is sufficiently accurate for understanding the subsequent evolution of the later Universe, it, of course, misses the dynamics of reheating. The duration of reheating depends on the underlying dynamics, and the effective equation of state may substantially deviate from the standard radiation-dominated case. For these reasons, we list below a few benchmark models of reheating whose equation of state spans the range $[0,1]$. Note that the selected models are for illustrative purposes, but our calculations can be applied to any other model, power spectrum shape, and type of primordial non-Gaussianity.   
\paragraph{Chaotic models ($w\sim 0.2,\, 0.25$)}
In these models, the inflaton $\phi$ is the dominant field during inflation with other fields ($\chi$,\dots) being subdominant~\cite{Linde:1990flp,Felder:2000hr}. As inflation ends, the inflaton field starts oscillating around the minimum of its potential, with large oscillations.

Since the frequency of these oscillations is higher than the expansion of the Universe and the potential is effectively quadratic, the effective averaged equation of state of the Universe coincides with a matter-dominated one, $w=0$. Due to coupling with the subdominant fields, the inflaton starts to decay. As shown by~\cite{Podolsky:2005bw}, the transition from this stage to an almost radiation-dominated Universe (preheating) is quite fast. However, since the inflaton has not fully decayed, it still gives a residual contribution to the energy of the Universe, leading the equation of state to be slightly lower than $1/3$. The speed of the sharp transition weakly depends on the coupling of the interaction, while the intermediate value of the equation of state varies in the range $0.2-0.3$~\cite{Podolsky:2005bw}. With reference to these models, we consider as benchmark cases $w=0.2,\, 0.25$.
\paragraph{Cannibal Dark Matter ($w\sim 0.1$)}
The so-called ``cannibalism'' happens when the lightest hidden-sector particle shows specific self-interactions which remain efficient even after the particle has become non-relativistic~\cite{Carlson:1992fn,Pappadopulo:2016pkp,Farina:2016llk,Erickcek:2020wzd}. In addition, this cannibal particle can dominate the Universe's energy density, leading to non-standard evolution histories. The equation of state of the Universe lies somewhere between radiation and matter, exactly due to the self-interaction process. Self-interactions heat the cannibal species, causing a mere logarithmic drop in its temperature with the scale factor. However, the cannibal particle can be unstable and finally decay into standard model particles, which will constitute the radiation bath. In these models, as shown in \cite{Erickcek:2020wzd}, both $w$ and $c_s^2$ reach an effective value of $0.1$ before the freeze-out of cannibal interactions. After that, the cannibal field evolves like pressureless matter and finally decays into standard model radiation when the Hubble rate is comparable to its decay width.
\paragraph{Power-law oscillations ($w=1/2$)} 
As argued by~\cite{Bernal:2019mhf,Garcia:2020eof, Garcia:2023dyf}, the specific shape of the inflaton potential around the minimum alters the subsequent reheating stage. For a generalised potential $V(\phi)\sim \phi^k$, the inflaton condensate energy density scales as $\sim a^{-6k/(k+2)}$ and the dynamics of the reheating stage depend on the value of $k$ considered, but also on the type of decay products (fermions or bosons). The equation of state can be generally written as $w=(k-2)/(k+2)$. When $k\geq 4$, scatterings can produce fluctuations in the inflaton which behave as a gas of massless particles, and the condensate slowly evolves from an equation of state $w>1/3$ to the standard radiation-dominated epoch, with the condensate fragmenting through the presence of oscillons. In this case, we consider the benchmark value $k=6$, which corresponds to $w=1/2$, whereas $k=2,\,4$ correspond to $w=0,\,1/3$, respectively.

\paragraph{Stiff-fluid case  ($w\sim 0.9,\,1$)} Lastly, we consider the case of a stiff fluid with $w\sim 1$, typical of quintessential inflation scenarios \cite{Spokoiny:1993kt,Peebles:1998qn,Brax:2005uf,Hossain:2014xha,Bettoni:2021qfs} (see also the Ricci reheating scenario \cite{Opferkuch:2019zbd,Laverda:2023uqv}). This case is of interest since the enhancement discussed around \cref{eq:frh_dep} may be largest. In particular, we fix with $w=c_s^2=0.9$, which would be the case of a universe close to kinetic domination~\cite{Spokoiny:1993kt}.  Note that, although no resonance is produced in the SIGW spectrum at exactly $c_s^2=1$ by momentum conservation, the resonance remains for a speed of sound relatively close to unity, although very sharp in the kernel \cite{Domenech:2019quo,Witkowski:2021raz}. 

\subsection{Spectra}
\label{sec:spectra_eval}
We explicitly report below the expression for the dimensionless SIGW spectral density in the Gaussian and non-Gaussian cases. The different spectra can be obtained by substituting \cref{eq:tensor_h_sigw_2} (or equivalently \cref{eq:tensor_h_sigw}) into \cref{eq:4pf} and evaluating the 4-point function in the Gaussian and non-Gaussian case. We refer to app.~\ref{app:Psh} and \cite{Perna:2024ehx} for explicit expressions of the tensor spectra up to the fifth order in the expansion.

%%%%%
\subsubsection{Gaussian Case}
For the Gaussian case, we have that
\bea
\Omega_{\rm GW}|_{\rm G}(k,\eta) 
    {=} & \frac{1}{48} x^2(1+b)^{-2}\frac{k^3}{2\pi^2} 2^5 \int \frac{d^3 q}{(2\pi)^3}\sum_{\lambda=+,\times} Q^2_{\lambda}(\vec{k},\vec{q})\overline{\mathbb{I}^2(|\vec{k}-\vec{q}|,q,\eta)} P_{\zeta_{\rm G}}(q)P_{\zeta_{\rm G}}(|\vec{k}-\vec{q}|)\,.
\eea
After introducing the usual change of variables, which amounts to
\be
\begin{cases}\label{eq:changeofvariables}
v = \frac{q}{k}\\
u = \frac{|\vec{k}-\vec{q}|}{k}
\end{cases}\quad \Rightarrow \quad \int d^3q_i = k^3 \int_0^\infty dv \int_{|1-v|}^{1+v}du \,u v \int_0^{2\pi}d\phi\,,
\ee
we obtain that the SIGW spectral density is given by
\bea
\label{Eq::Omega_G}
    \Omega_{\rm GW}(k,\eta)|_{\rm G} 
    {=} & x^{-2b}\frac{4^{2b}(1+b)^{-2}}{3c_s^4}\left(\frac{b+2}{3+2b}\right)^2\Gamma^4\left[b+\frac{3}{2}\right] \\
    & \times \int_0^\infty dv \int_{|1-v|}^{1+v} {du}\left(\frac{4v^2-(1+v^2-u^2)^2}{4v^2u^2}\right)^2 |1-y^2|^b \mathcal{P}_{\zeta_{\rm G}}(kv)\mathcal{P}_{\zeta_{\rm G}}(ku)\\
    & \times\left\{ I_A^2(y_1,u_1,v_1,c_s,b)+ \frac{4}{\pi^2} I_B^2(y_1,u_1,v_1,c_s,b)\right\}\,,
\eea
where $\mathcal{P}_{\zeta_{\rm G}}\equiv \frac{k^3}{2\pi^2}P_{\zeta_{\rm G}}$ is the dimensionless curvature power spectrum.
%%%%%
\subsubsection{Non-Gaussian cases}

Primordial NG contributions leave a specific imprint on the frequency-dependent part, and we can distinguish three different non-vanishing contributions, which read \cite{Adshead:2021hnm,Garcia-Saenz:2022tzu,Perna:2024ehx}
\bea
\label{eq:t_contrib}
    \Omega_{\rm GW}(k,\eta)|_{\rm t} 
    {=} & \frac{1}{48} x^2(1+b)^{-2}\frac{k^3}{2\pi^2} 2^8 \left( \frac{3}{5}f_{\rm NL}\right)^2 \int \frac{d^3 q_1}{(2\pi)^3} \int \frac{d^3 q_2}{(2\pi)^3}\sum_\lambda Q_{\lambda}(\vec{k},\vec{q}_1)Q_{\lambda}(\vec{k},\vec{q}_2)\\
    & \hspace{3.5cm}\times \overline{\mathbb{I}(|\vec{k}-\vec{q}_1|,q_1,\eta)\mathbb{I}(|\vec{k}-\vec{q}_2|,q_2,\eta)}\\
    &\hspace{3.5cm}\times  \mathcal{P}_{\zeta_{\rm G}}(q_2)\mathcal{P}_{\zeta_{\rm G}}(|\vec{k}-\vec{q}_2|)\mathcal{P}_{\zeta_{\rm G}}(|\vec{q}_1-\vec{q}_2|)\,,
\eea
\bea
\label{eq:u_contrib}
    \Omega_{\rm GW}(k,\eta)|_{\rm u}
    {=} & \frac{1}{48} x^2(1+b)^{-2}\frac{k^3}{2\pi^2} 2^8 \left( \frac{3}{5}f_{\rm NL}\right)^2 \int \frac{d^3 q_1}{(2\pi)^3} \int \frac{d^3 q_2}{(2\pi)^3}\sum_\lambda Q_{\lambda}(\vec{k},\vec{q}_1)Q_{\lambda}(\vec{k},\vec{q}_2)\\
    & \hspace{3.5cm}\times \overline{\mathbb{I}(|\vec{k}-\vec{q}_1|,q_1,\eta)\mathbb{I}(|\vec{k}-\vec{q}_2|,q_2,\eta)}\\
    &\hspace{3.5cm}\times \mathcal{P}_{\zeta_{\rm G}}(q_1)\mathcal{P}_{\zeta_{\rm G}}(q_2)\mathcal{P}_{\zeta_{\rm G}}(|\vec{k}-(\vec{q}_1+\vec{q}_2)|)\,,
\eea
and
\bea
\label{eq:hybrid_contrib}
    \Omega_{\rm GW}(k,\eta)|_{\rm hybrid}
    {=} & \frac{1}{48} x^2(1+b)^{-2}\frac{k^3}{2\pi^2} 2^7 \left( \frac{3}{5}f_{\rm NL}\right)^2 \int \frac{d^3 q_1}{(2\pi)^3}\int \frac{d^3 q_2}{(2\pi)^3}\sum_\lambda Q^2_{\lambda}(\vec{k},\vec{q}_1)\\
    & \times \overline{\mathbb{I}^2(|\vec{k}-\vec{q}_1|,q_1,\eta)} \mathcal{P}_{\zeta_{\rm G}}(|\vec{k}-\vec{q}_1|)\mathcal{P}_{\zeta_{\rm G}}(q_2)P_{\zeta_{\rm G}}(|\vec{q}_1-\vec{q}_2|)\,,
\eea
where we followed the notation of \cite{Garcia-Saenz:2022tzu,Perna:2024ehx}.
After the change of variables\footnote{Note that in the connected terms we can adopt similar changes of variable as in the Gaussian case, while the disconnected one require a slightly different change of variables. We refer to~\cite{Adshead:2021hnm,Perna:2024ehx} for more details.} given in \cref{eq:changeofvariables}, we find that they respectively become
\bea
    \Omega_{\rm GW}(k,\eta)|_{\rm t} 
    {=} & \frac{2}{3\pi}(1+b)^{-2} x^{-2b} \left(\frac{b+2}{3+2b}\right)^2 \Gamma^4\left[b+\frac{3}{2}\right]\frac{4^{2b}}{c_s^4}\left( \frac{3}{5}f_{\rm NL}\right)^2\\
    & \times \int_0^\infty dv_1 \int_0^\infty dv_2 \int_{|1-v_1|}^{1+v_1}du_1\int_0^{2\pi} d\phi_{12}\int_{|1-v_2|}^{1+v_2}du_2\, {v_1^2v_2^2u_1^2u_2^2}\\
    & \times \left(\frac{4v_1^2-(1+v_1^2-u_1^2)^2}{4v_1^2}\right)\left(\frac{4v_2^2-(1+v_2^2-u_2^2)^2}{4v_2^2}\right){|1-y_1^2|^{b/2}}{|1-y_2^2|^{b/2}}\\
    &\times\left\{ I_A(y_1,u_1,v_1,c_s,b)I_A(y_2,u_2,v_2,c_s,b)+ \frac{4}{\pi^2} I_B(y_1,u_1,v_1,c_s,b)I_B(y_2,u_2,v_2,c_s,b)\right\}\\
& \times \frac{\mathcal{P}_{\zeta_{\rm G}}(kv_2)}{v_2^3}\frac{\mathcal{P}_{\zeta_{\rm G}}(ku_2)}{u_2^3}\frac{\mathcal{P}_{\zeta_{\rm G}}(kw_{12a})}{w_{12a}^3}\,,
\eea
and
\bea
    \Omega_{\rm GW}(k,\eta)|_{\rm u} 
    {=} & \frac{2}{3\pi}(1+b)^{-2} x^{-2b} \left(\frac{b+2}{3+2b}\right)^2 \Gamma^4\left[b+\frac{3}{2}\right]\frac{4^{2b}}{c_s^4}\left( \frac{3}{5}f_{\rm NL}\right)^2\\
    & \times \int_0^\infty dv_1 \int_0^\infty dv_2 \int_{|1-v_1|}^{1+v_1}du_1\int_0^{2\pi} d\phi_{12}\int_{|1-v_2|}^{1+v_2}du_2\, {v_1^2v_2^2u_1^2u_2^2}\\
    & \times \left(\frac{4v_1^2-(1+v_1^2-u_1^2)^2}{4v_1^2}\right)\left(\frac{4v_2^2-(1+v_2^2-u_2^2)^2}{4v_2^2}\right){|1-y_1^2|^{b/2}}{|1-y_2^2|^{b/2}}\\
    &\times\left\{ I_A(y_1,u_1,v_1,c_s,b)I_A(y_2,u_2,v_2,c_s,b)+ \frac{4}{\pi^2} I_B(y_1,u_1,v_1,c_s,b)I_B(y_2,u_2,v_2,c_s,b)\right\}\\
& \times \frac{\mathcal{P}_{\zeta_{\rm G}}(kv_1)}{v_1^3}\frac{\mathcal{P}_{\zeta_{\rm G}}(kv_2)}{v_2^3}\frac{\mathcal{P}_{\zeta_{\rm G}}(kw_{12b})}{w_{12b}^3}\,,
\eea
%%%%
where we introduced $w_{12a}=|\mathbf{q_1}-\mathbf{q_2}|/k$, $w_{12b}=|\mathbf{k}-\mathbf{q_1}-\mathbf{q_2}|/k$, and $\phi_{12}=\phi_1-\phi_2$. Lastly, we also have that
%%%%
\bea
\Omega_{\rm GW}(k,\eta)|_{\rm hyb} = & \frac{2}{3}(1+b)^{-2} x^{-2b} \left(\frac{b+2}{3+2b}\right)^2 \Gamma^4\left[b+\frac{3}{2}\right]\frac{4^{2b}}{c_s^4}\left( \frac{3}{5}f_{\rm NL}\right)^2\\
& \times \int_0^\infty dv_1\int_{|1-v_1|}^{1+v_1}du_1\int_0^\infty dv_2\int_{|1-v_2|}^{1+v_2}du_2 \frac{v_1^6u_2v_2}{u_1}\\
& \times \left(\frac{4v_1^2-(1+v_1^2-u_1^2)^2}{4v_1^2}\right)^2|1-y_1^2|^{b}\\
& \times\left\{ I_A^2(y_1,u_1,v_1,c_s,b)+ \frac{4}{\pi^2} I_B^2(y_1,u_1,v_1,c_s,b)\right\}\\
& \times \frac{\mathcal{P}_{\zeta_{\rm G}}(ku_1)}{u_1^3}\frac{\mathcal{P}_{\zeta_{\rm G}}(kv_1v_2)}{v_1^3v_2^3}\frac{\mathcal{P}_{\zeta_{\rm G}}(kv_1u_2)}{v_1^3u_2^3}\,.
\eea
\begin{figure}
    \centering
    \includegraphics[width=1\linewidth]{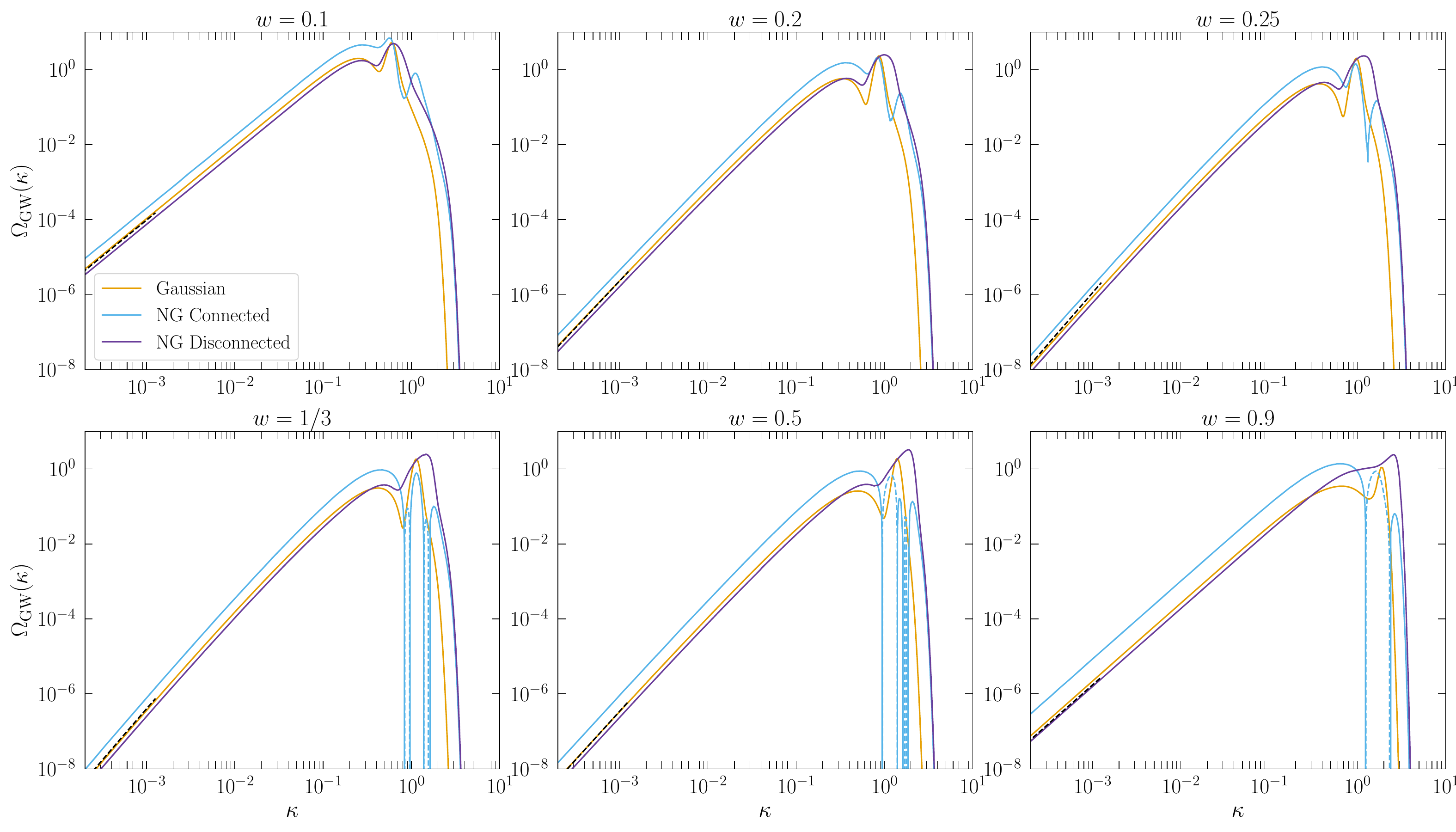}
    \caption{The figure shows the SIGWs spectrum for different values of the speed of sound and the equation of state obtained for a lognormal scalar power spectrum}. For each case, we show the Gaussian, non-Gaussian connected, and non-Gaussian disconnected spectra, in orange, blue, and purple solid lines, respectively. A dashed blue line signals that the non-Gaussian connected piece becomes negative. The dashed black line indicates the IR analytical behaviour from \eqref{eq:IR_behav}. The Gaussian (non-Gaussian) spectra are normalised by $A^2$ ($A^2 (3/5)^2 f_{\rm NL}^2$). With reference to \cref{eq:frh_dep}, and to remove any dependence on the reheating scale, all spectra are further normalised over $(k_*/k_{\rm RH})^{-2b}$.
    \label{fig:spectra}
\end{figure}

We report our results in \cref{fig:spectra}. For simplicity, we adopt the usual lognormal ansatz for the scalar power spectrum in evaluating the SIGW spectra typical of some multifield models~\cite{Garcia-Bellido:1996mdl, Kawasaki:2015ppx, Pi:2017gih, Fumagalli:2020nvq, Fumagalli:2021cel, Kallosh:2022vha, Braglia:2022phb, Tada:2023pue, Tada:2023fvd,Wang:2024vfv,Iacconi:2024hmg}, namely
\begin{equation}
    \mathcal{P}^{\rm LN}(k) =\frac{A}{\sqrt{2\pi \sigma^2}} \exp\left[-\frac{1}{2\sigma^2} \log^2 \left(\frac{k}{k_{*}}\right)\right]\,.
    \label{eq:ans_lognormal}
\end{equation}
Each panel shows the Gaussian, the sum of the non-Gaussian connected spectra (u+t), and the non-Gaussian disconnected contribution, respectively, normalised by $A^2$ and $A^3 (3/5f_{\rm NL})^2$. To verify our numerical results, we report that the IR behaviour coincides with the Gaussian case for $k/k_*\ll1$. As explained in~\cite{Domenech:2021ztg}, the IR analytical behaviour can be obtained starting from the results for a Dirac delta spectrum and smoothing them, following the procedure of~\cite{Pi:2020otn}. For a Dirac delta spectrum, one obtains
\be
\label{eq:IR_behav}
\Omega_{\rm GW}^{\delta}(\kappa\ll1,b,c_s^2)=
\begin{cases}

\displaystyle
\frac{1}{12}\left[\frac{2^{1+b}(2+b)\,\Gamma\!\left(\frac{3}{2}+b\right)^2}{\pi\, (c_s^2)^{1+b}\,(1+b)^{1+b}}\right]^2\left[\frac{\pi}{\sin(\pi b)\,\Gamma(2+b)} \right]^2\, x^{2+2b}& b<0 \\[1.2em]

\displaystyle
\frac{1}{24\pi}\left[\frac{(2+b)(1+b+b^2)}{c_s^2\, b\,(1+b)^2}\right]^2\frac{1}{2}\left[\frac{2^{1+b}}{(1+b)^{1+b}}\Gamma\!\left(b+\frac{3}{2}\right)\right]^2\, x^{2-2b}& b>0 \\[1.2em]

\displaystyle
\frac{x^2}{3\,c_s^4}\,\log^2\!\left(\frac{1}{x}\right)& b=0
\end{cases}
\ee
and the results for the log-normal spectrum can be obtained as\footnote{See also app. A of~\cite{Dandoy:2023jot} for corrections in the IR tail of very narrow spectra.}
\be
\Omega_{\rm GW}^{\rm LN}(\kappa) = {\rm erf}\left(\frac{\kappa}{2\sigma}\right)\Omega_{\rm GW}^{\delta}(\kappa)\,.
\ee

\begin{figure}
    \centering
    \begin{minipage}{0.49\linewidth}
        \centering
        \includegraphics[width=\linewidth]{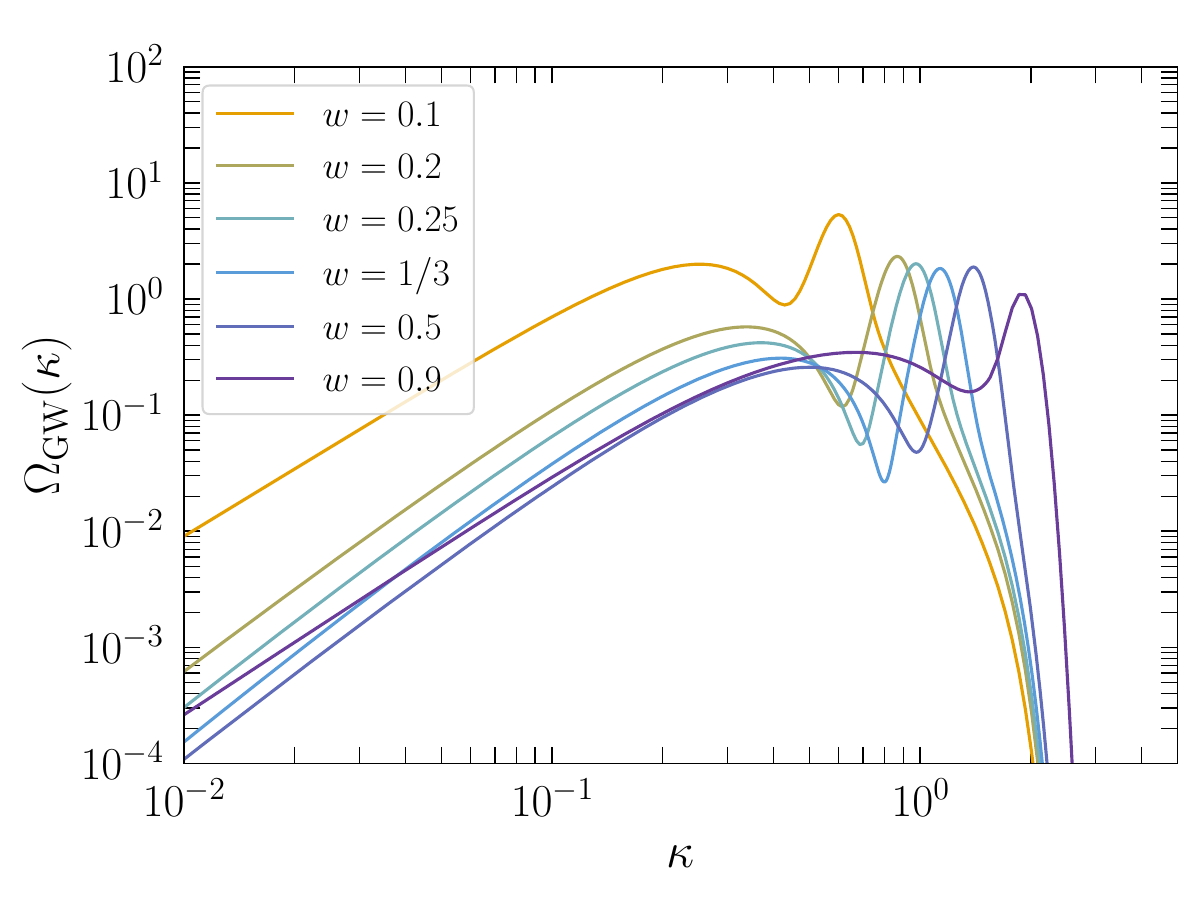}
    \end{minipage}
    \hfill
    \begin{minipage}{0.49\linewidth}
        \centering
        \includegraphics[width=\linewidth]{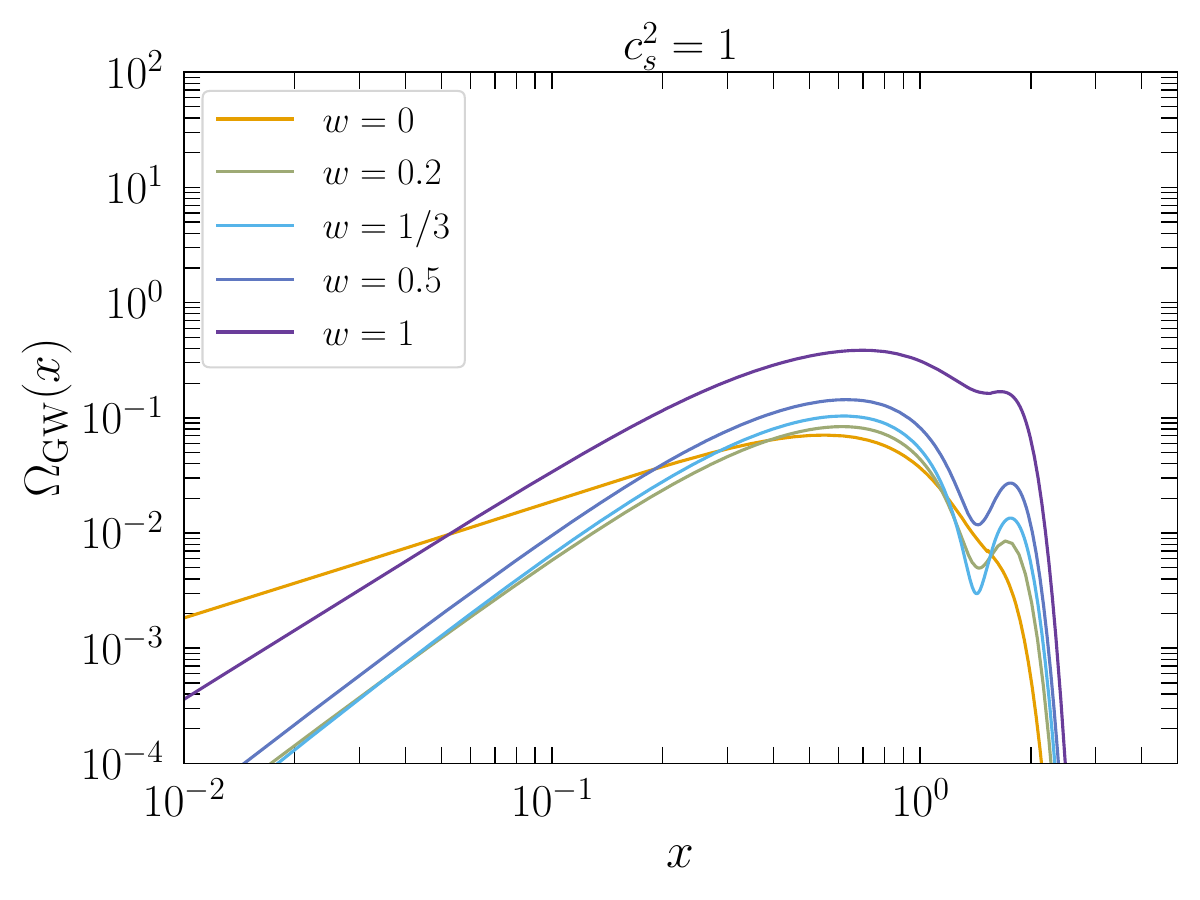}
    \end{minipage}
    \caption{\textbf{Left Panel:} Comparison of the shape of the Gaussian spectrum obtained for different values of $w=c_s^2$ for a lognormal scalar power spectrum. All the spectra are normalised as in \cref{fig:spectra}. \textbf{Right Panel:} Comparison of different spectra at fixed $c_s^2=1$ and for different values of $w$ in the case of a lognormal scalar power spectrum.}
    \label{fig:Gaussian_Comparison}
\end{figure}

In \cref{fig:spectra}, we also see how all the spectra exhibit the resonant peak typical of SIGW signals when $c_s^2\neq 1$. Interestingly, we observe that the peak is well defined for all values of $w$ considered, as expected, and begins to smooth out only for $w=0.9$. In this case, as we approach $c_s^2=1$, the resonance gradually disappears, as argued above. When primordial NG is included, additional peaks arise, generalising the results already found in earlier works~\cite{Adshead:2021hnm,Perna:2024ehx}. Also in this case, as $c_s^2\to1$, the resonance is suppressed, and the peak becomes smoother. An example of the smoothing can be seen in the left panel of \cref{fig:Gaussian_Comparison} in the Gaussian case and in all the curves in the right panel as well.

It should be noted that some of the non-Gaussian connected contributions take negative values in certain frequency ranges. This can be understood from the generalised kernel, \cref{eq:generalized_kernel}, which can become negative for specific momentum configurations (see the minus sign associated with $\mathcal{Q}$). However, it is important to stress that the final GW energy density must be positive, since by definition $\Omega_{\rm GW}\sim h_{ij}^2$\footnote{We specify that the final total GW spectrum being negative (even if only in some frequency ranges) would indicate a violation of the perturbativity condition, making the final calculation unreliable in those frequency ranges. In this work, we checked that the spectra are positive in the whole range considered.}. Nevertheless, the presence of specific features that depend on the values of $w$ and $c_s^2$ suggests that, if observed, these signals could provide direct information about the evolution of the epoch during which they were produced. In fact, all regions of the spectrum are relevant for this purpose, as we show in both the panels of~\cref{fig:Gaussian_Comparison}. In particular, in the left panel we plot the Gaussian spectra obtained for different values of $w=c_s^2$ to show how their features vary as the spectrum changes, while in the right panel we report an illustrative example, fixing $c_s^2=1$ and varying $w$, to show that not only the effects of different $w$ are still observable when the speed of sound is fixed, but the spectra still show specific features even in the case in which the resonant peak is absent. 

Below, we list the main distinct imprints:

\begin{itemize}
\item[(i)] The slope of the low-frequency, infrared (IR) tail strongly depends on the value of $w$ (or equivalently $b$), as discussed above (see also \cite{Domenech:2020kqm}), while it is independent of the speed of sound, which instead affects only the overall amplitude. Therefore, detecting only the IR tail would already provide a clear indication of the equation of state during reheating.

\item[(ii)] The central features around the peak, although qualitatively similar among the different spectra, vary as $w$ and $c_s^2$ change. In particular, increasing $c_s^2$ shifts the peak toward higher frequencies \cite{Domenech:2019quo}. This behaviour can be understood from the resonance condition, which depends on the relation between the sound horizon and the Hubble horizon: a larger sound speed implies that perturbations propagate more efficiently, effectively modifying the characteristic scale at which the resonance occurs and thus moving the peak to higher $k$ (or $f$).

\item[(iii)] The steepness of the decrease immediately after the peak, which defines the UV tail, slightly changes as the equation of state is varied (see also \cite{Atal:2021jyo}).
\end{itemize}

\subsection{Fisher}
To assess LISA’s capability to detect the SIGW signal originating from different reheating epochs, we perform a Fisher forecast to predict the expected uncertainty of a given experiment in measuring the model parameters. The Fisher matrix is defined as
\begin{equation}
     F_{\alpha\beta} = \left\langle-\frac{\partial^2 \ln \mathcal{L}}{\partial q_{\alpha}\partial q_{\beta}} \right\rangle \bigg|_{\vec{q} = \vec{q}_0},
\end{equation}
with $q_\alpha$ the parameters of interest, $\vec{q}_0$ their best-fit values and $\mathcal{L}$ the likelihood. We consider a Gaussian likelihood defined as
\begin{equation}
    \ln \mathcal{L} =-\frac{N_c}{2}\sum_{i}\sum_{k}\frac{\left(\mathcal{D}_{i}^{(k)}-\mathcal{D}_{i}^{(k), \rm{th}}\right)^2}{\sigma_{i}^{(k) 2}}\,,
\end{equation}
where $i$ runs over the three LISA (TDI) channels and $k$ runs over frequency bins $f_k$.  $N_c$ is the number of data segments in the analysis. The variance can be expressed in terms of the theoretical ansatz as $\sigma_{i}^{(k) 2} = \left(\mathcal{D}_{i}^{(k), \rm{th}}\right)^2$ and $\mathcal{D}_{i}^{(k), \rm{th}} (f_k, \vec{\lambda}) = \mathcal{R}_{ii}(f_k)h^2 \Omega_{\rm GW}(f_k, \vec{\lambda}) + N^{\Omega}_{ii}(f_k)$ with $\mathcal{R}_{ij}(f_k)$ the response function and $N^{\Omega}_{ij}(f_k)$ the noise associated to the TDI channels (see e.g., ~\cite{LISACosmologyWorkingGroup:2024hsc} for more details). Lastly, we obtain

\begin{equation}
    F_{\alpha\beta} = T_{\rm obs} \sum_{i \in \{A,E,T\}} \int_{f_{\rm min}}^{f_{\rm max}} \frac{\partial \mathcal{D}_{i}^{\rm{th}}}{\partial q_\alpha} \frac{\partial \mathcal{D}_{i}^{\rm{th}}}{\partial q_\beta} \frac{1}{\sigma_i^{2}}df
\end{equation}
where $T_d$ is the total observation time and $\sigma$ accounts for the noise in all three TDI channels. After all the entries of the Fisher matrix are evaluated, its inverse returns the covariance matrix. In this section, we assume the signal to be centred in the LISA band and analyse the effect of a different peak frequency in the next section. In what follows, we consider an observation time of $4$ years.

\begin{figure}
    \centering
    \includegraphics[width=\linewidth]{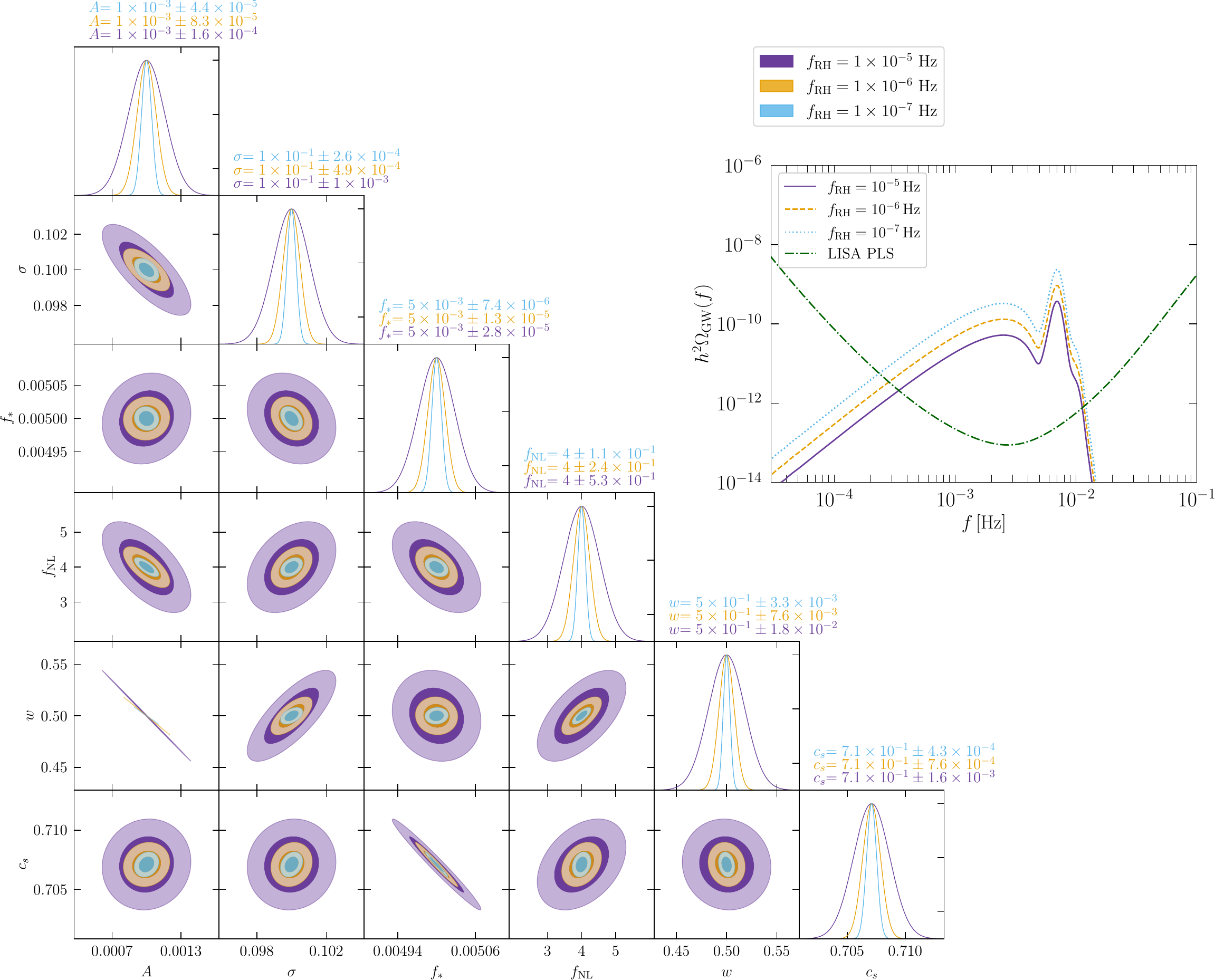}
    \caption{\textbf{Left Panel:} Contour plot showing the Fisher forecast in the case $w=0.5$. Different colours correspond to different values of $f_{\rm RH}$. \textbf{Right Panel:} Spectra corresponding to the benchmark values adopted for the Fisher. We also report the power-law-integrated sensitivity for LISA, obtained using \texttt{GWBird}~\cite{Caporali:2025dyf}, for 4 years of observation time and an SNR$=10$, typically used to assess a detection.}
    \label{fig:fisher_0.5}
\end{figure}

\begin{figure}
    \centering
    \includegraphics[width=\linewidth]{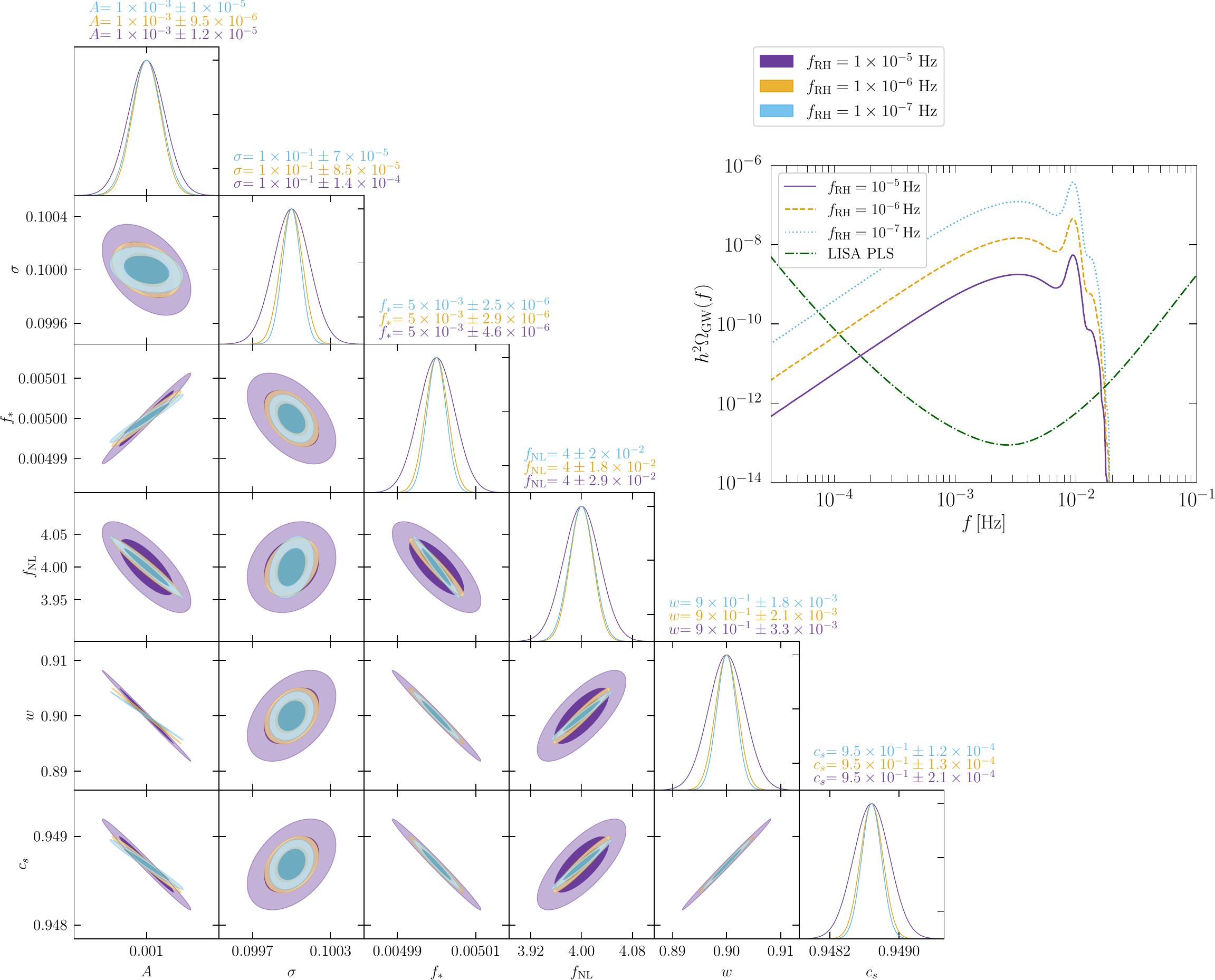}
 \caption{\textbf{Left Panel:} Contour plot showing the Fisher forecast in the case $w=0.9$. Different colours correspond to different values of $f_{\rm RH}$. \textbf{Right Panel:} Spectra corresponding to the benchmark values adopted for the Fisher. We also report the power-law-integrated sensitivity for LISA, obtained with \texttt{GWBird}~\cite{Caporali:2025dyf}, for 4 years of observation time and an SNR$=10$, typically used to assess a detection.}
    \label{fig:fisher_0.9}
\end{figure}

Previous works~\cite{Perna:2024ehx,LISACosmologyWorkingGroup:2025vdz} showed that shape parameters such as the peak frequency, roughly given by $f_*=k_*/2\pi$, and the width $\sigma$ can be reconstructed with very high precision, thanks to the distinctive imprints they leave on the spectrum. In particular, the resonance peak is the most characteristic feature of SIGW signals; therefore, its detection, or even its non-detection, directly provides information about the peak frequency. A similar argument applies to $\sigma$, which determines the width of both the resonance peak and the secondary peak. This can also be seen in~\cite {Pi:2020otn}, where the authors derived analytical templates around the peak and studied its behaviour both near the peak and in the tails. We find similar results in our analysis, obtaining errors of the per-mille or smaller on this parameters when the signal is perfectly detected (see e.g. \cref{fig:fisher_0.5} or \cref{fig:fisher_0.9}).

In addition, as we argued in~\cref{sec:Derivation}, the reheating frequency $f_{\rm RH}$ plays a crucial role in detectability and has a non-negligible impact on the results. However, being a pure multiplicative parameter that appears from \cref{eq:frh_dep}, it is fully degenerate with the amplitude and, if included, it would lead to a singular matrix. To account for the effect of different reheating temperatures, we repeat the analysis for different benchmark values of $f_{\rm RH}$ (see \cref{eq:krh} for the relation between $f_{\rm RH}$ and $T_{\rm RH}$). % In choosing the reheating temperature,
Note that we always consider $ f_{\rm RH}\leq f^{{\rm min}}_{\rm LISA}< f_*$, where $f^{{\rm min}}_{\rm LISA}=10^{-5}\,{\rm Hz}$, to be consistent with the analytical kernel. Such requirements come from simplicity. First, if $f_*<f_{\rm RH}$, the SIGWs are generated during the radiation domination era, which has already been analysed in \cite{Unal:2018yaa, Adshead:2021hnm, Perna:2024ehx}. Second, if $f_{\rm RH}>f^{{\rm min}}_{\rm LISA}$, and the reheating scale enters the LISA band, the SIGW presents additional features around $f\sim f_{\rm RH}$. 

We omit this possibility as the precise feature depends on the transition to radiation domination, and an exact analytical kernel is currently unavailable. In our case of interest, that is for $ f_{\rm RH}\leq f^{{\rm min}}_{\rm LISA}< f_*$, our SIGW spectrum is expected to be insensitive to the details of the transition as long as it takes less than an e-fold \cite{Domenech:2020kqm}.

\begin{figure}
    \centering
    \includegraphics[width=\linewidth]{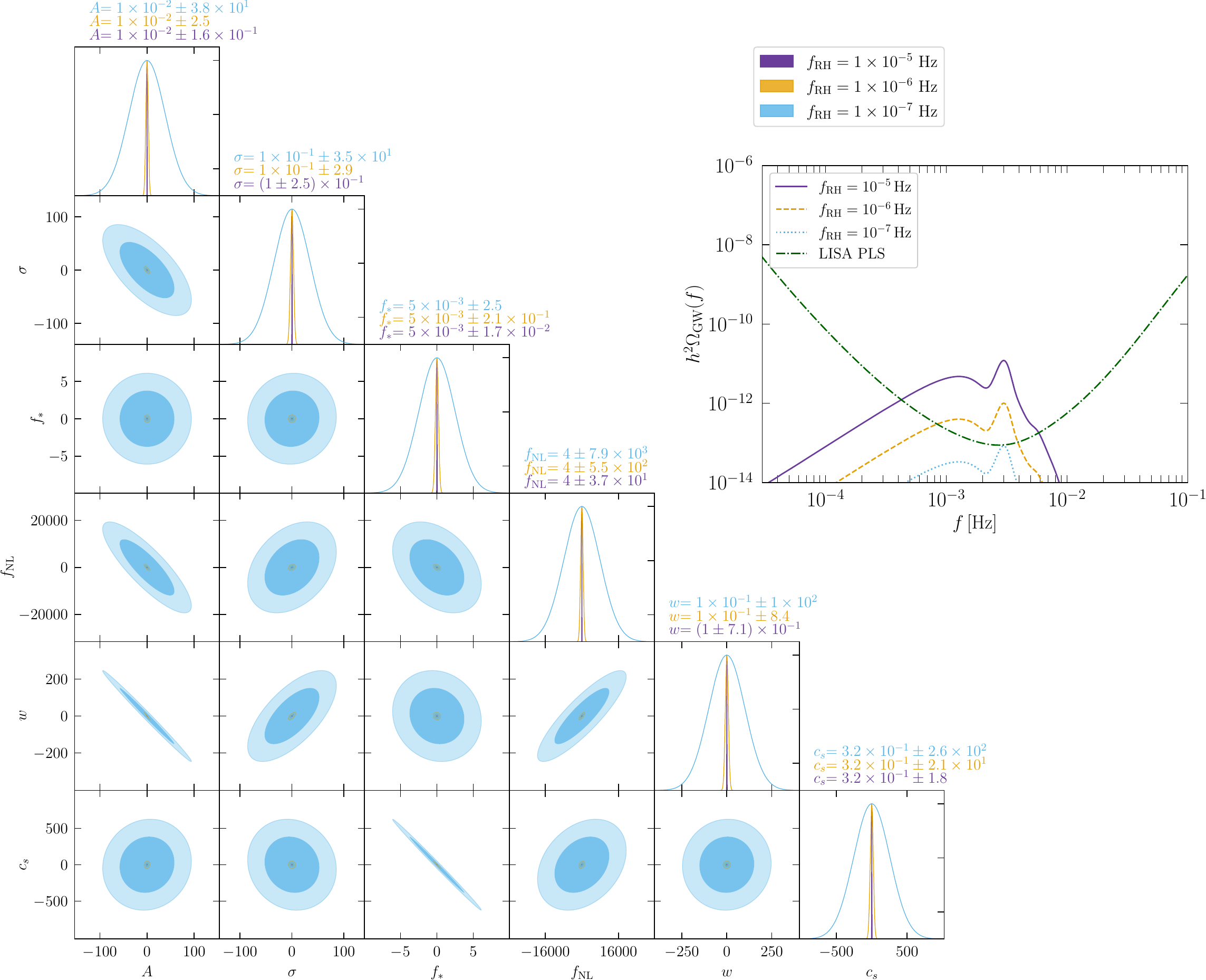}
\caption{\textbf{Left Panel:} Contour plot showing the Fisher forecast in the case $w=0.1$. Different colours correspond to different values of $f_{\rm RH}$. \textbf{Right Panel:} Spectra corresponding to the benchmark values adopted for the Fisher. We also report the power-law-integrated sensitivity for LISA, obtained with \texttt{GWBird}~\cite{Caporali:2025dyf}, for 4 years of observation time and an SNR$=10$, typically used to assess a detection.}
    \label{fig:fisher_0.1}
\end{figure}

We repeat the analysis for three different values of $w$ and $c_s^2=w$: respectively $0.1$ for the cannibal dark matter (similar results hold for $w=0.2$ as well), $0.5$ for the power-law oscillations, and $0.9$ for the stiff-fluid case. The results are reported respectively in \cref{fig:fisher_0.1,fig:fisher_0.5,fig:fisher_0.9}. In the left panels, we report the outcomes of the Fisher forecast, and in the right panel, the corresponding spectra. To ease understanding of the results, we also plot the integrated power-law sensitivity for LISA over 4 years, with a signal-to-noise ratio threshold ${\rm SNR}_{\rm thr} = 10$. We adopt the same values in the Fisher analysis as well. We proceed to discuss the cases $w=0.1$ and $w=\{0.5,\,0.9\}$ separately, as they qualitatively different, respectively corresponding to $w<1/3$ ($b>0$) and $w>1/3$ ($b<0$).

First, for $w=0.1$, we see from \cref{fig:fisher_0.1} that the constraining power of LISA improves as the reheating frequency increases. This is evident from \cref{eq:frh_dep}: when $b>0$, the spectrum is suppressed as the reheating temperature decreases. Thus, a higher amplitude is needed to detect it. For instance, as shown in the right panel of \cref{fig:fisher_0.1}, the SIGW spectra are strongly suppressed even considering an amplitude of $10^{-2}$, and could also not be observed after 4 years of observation time. 
Indeed, in the left panel of \cref{fig:fisher_0.1}, the corner plot shows that all the parameters are poorly reconstructed, even in the best case. It is already possible to note some interesting features. All the parameters exhibit a certain degree of degeneracy, which can be understood in light of the IR behaviour. All parameters contribute to the final amplitude of the spectrum, and the degeneracy is broken only due to specific dependencies inside the integrals, which impact the shape of the resonant and second peak, as well as the minimum between the two. The direction of the contours can be understood in a similar way. As \cref{eq:IR_behav} shows, $\Omega_{\rm GW}\propto A^2/c_s^4$. Thus, once a signal is detected (and the ratio $A^2/c_s^4$ is fixed), higher amplitudes correspond to higher speeds of sound to keep the ratio constant. The contour showing the correlation between the amplitude and $f_{\rm NL}$ has the opposite direction. This is because both of them appear as $Af_{\rm NL}^2$, and an increase of one parameter necessarily implies a decrease of the other. The degeneracy in this case is broken by the Gaussian term, which depends only on the amplitude and shows different spectral imprints with respect to the non-Gaussian ones, as shown in \cref{fig:spectra}. Similarly, the degeneracy between $f_*$ and $c_s$ can be understood by considering that both parameters control the position of the peak. The direction of the contour follows from the fact that increasing either parameter shifts the peak towards higher frequencies. Therefore an increase in $f_*$ necessarily has to be compensated by a decrease in $c_s$ to preserve the same peak position. We add that a more detailed analysis would require the use of priors to exclude negative values of the speed of sound or the amplitude, as well as values larger than unity. We omit the priors because we are interested in assessing the signal-detection capability. Such large uncertainties indicate the poor constraining power of LISA for signals produced during epochs with $w<1/3$. 

% \begin{figure}%[htbp]
%     \centering

%     \begin{minipage}[c]{0.59\textwidth}
%         \centering
%         \includegraphics[width=\textwidth]{Figures/Fisher_w_0.1_cs2_0.1_Full.pdf}
%     \end{minipage}
%     \hfill
%     \begin{minipage}[c]{0.4\textwidth}
%         \centering
%         \includegraphics[width=\textwidth]{Figures/Spectra_w_0.1_cs2_0.1.pdf}
%     \end{minipage}

%     \caption{\textbf{Left Panel:} Contour plot showing the Fisher forecast in the case $w=0.1$. Different colours correspond to different values of $f_{\rm RH}$. \textbf{Right Panel:} Spectra corresponding to the benchmark values adopted for the Fisher. We also report the power-law-integrated sensitivity for LISA, obtained with \texttt{GWBird}~\cite{Caporali:2025dyf}, for 4 years of observation time and an SNR$=10$, typically used to assess a detection.}
%     \label{fig:fisher_0.1}
% \end{figure}

The situation changes when $w>1/3$. We observe that, for both $w=0.5$ and $w=0.9$, the constraints increase as the reheating frequency decreases. Again, this follows from \cref{eq:frh_dep}, which justifies the boost of the power spectrum when $b<0$. To show its importance, we choose an amplitude of one order of magnitude smaller than in the previous case, that is $A=10^{-3}$. Nevertheless, in all cases we are able to perfectly reconstruct all the parameters, with a precision that ranges from almost 1\% in the worst case to $1/10^4$ in the best cases. In particular, as $w$ increases, the boost becomes more efficient at lower reheating frequencies, thereby improving the constraints.
The improved constraints on $f_{\rm NL}$ can be understood considering that the main imprint can be observed in the presence of a third peak arising in the UV tail, but also for a slight smoothing of the peak (see also \cite{Perna:2024ehx}). The enhancement allows these features to be detectable even when the amplitude is small. Note that the enhanced signal cannot exceed the BBN bounds on additional relativistic species \cite{Binetruy:2012ze,Caprini:2018mtu}. We accounted for this when choosing the benchmark values for this subsection and in the next.
Interestingly, when $w=0.9$ we notice that additional degeneracies arise. This behaviour can be understood by considering that as we approach the stiff matter limit with $c_s^2=1$, the resonance is suppressed and the characteristic resonant peak of the spectrum is smoothed out. The loss of these sharp structural features leaves more freedom to parameters to conspire leading to different configurations that mimick the observed spectrum: for example lateral shifts driven by $f_*$ can be closely mimicked by adjustments in parameters such as $A$, $w$, and $c_s$. Consequently, this increases their covariances and broadens the parameter constraints within the Fisher analysis.


\subsection{Scans}
\begin{figure}%[htbp]
    \centering

    % prima riga
    \begin{minipage}[c]{0.45\textwidth}
        \centering
        \includegraphics[width=\textwidth]{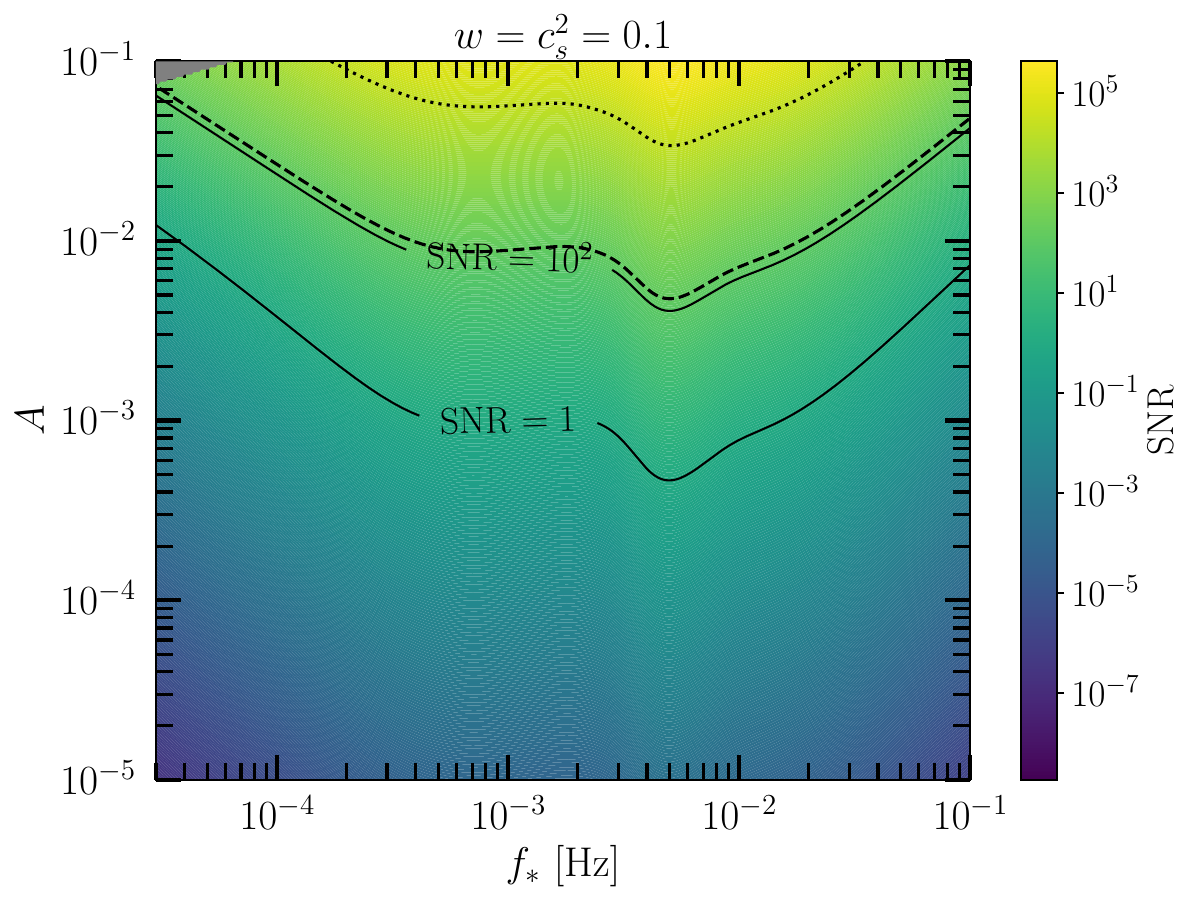}
    \end{minipage}
    \hfill
    \begin{minipage}[c]{0.45\textwidth}
        \centering
        \includegraphics[width=\textwidth]{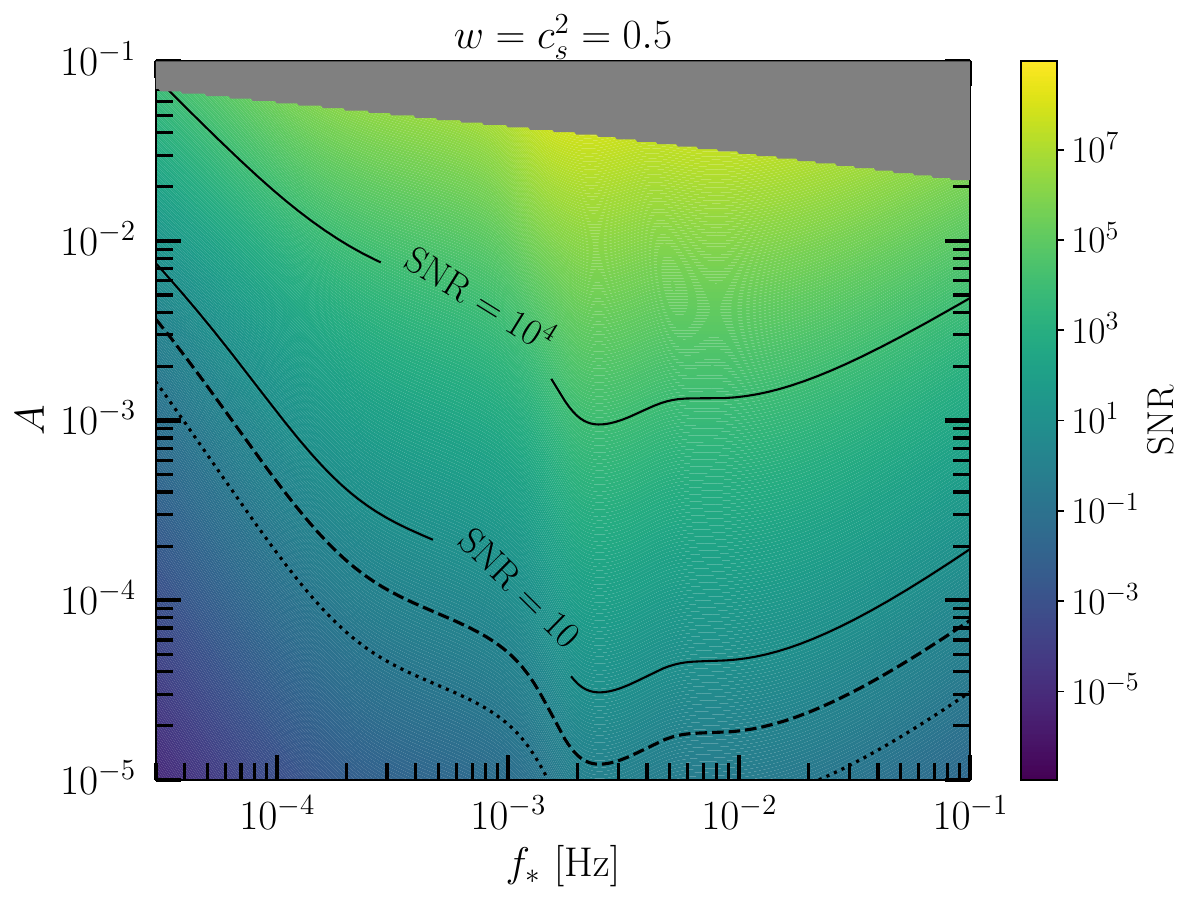}
    \end{minipage}

    \vspace{0.5cm}

    % seconda riga
    \begin{minipage}[c]{0.45\textwidth}
        \centering
        \includegraphics[width=\textwidth]{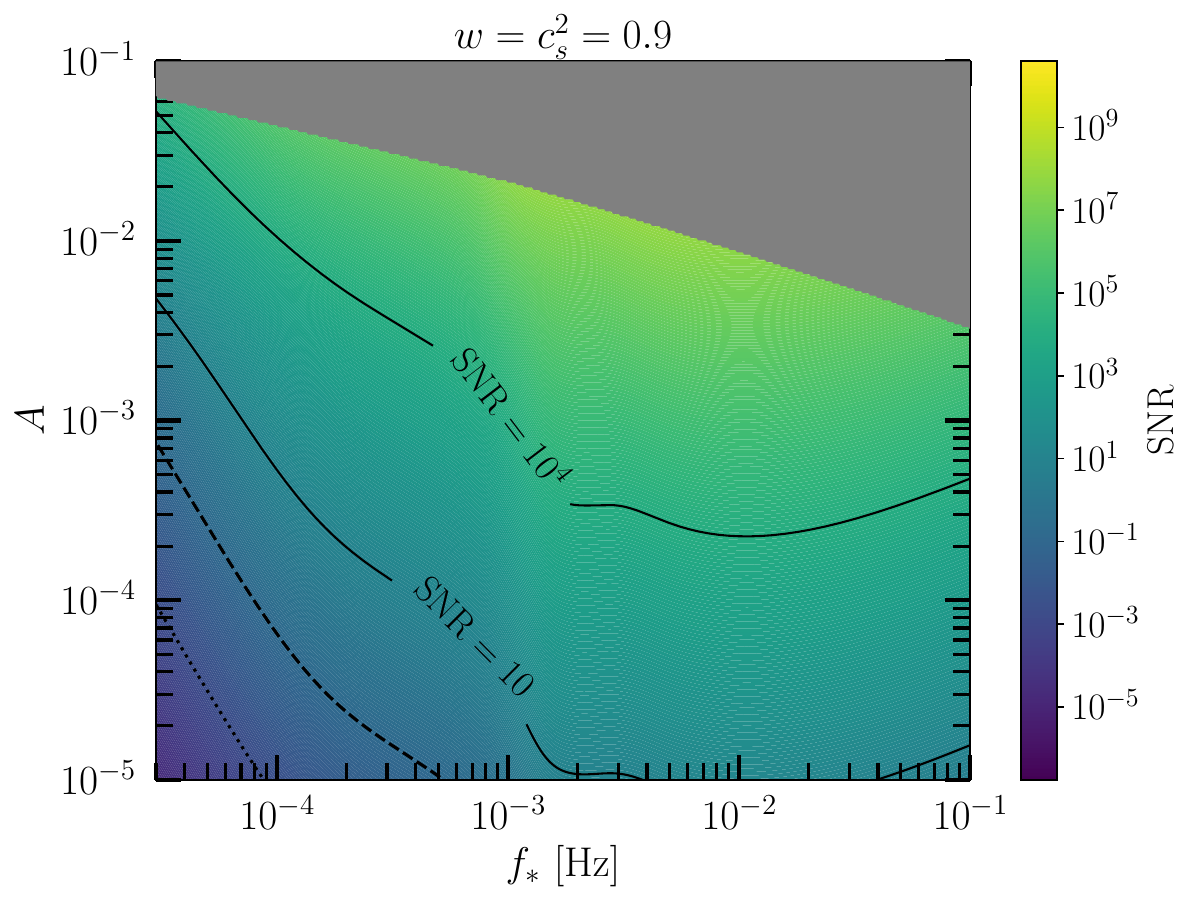}
    \end{minipage}

    \caption{Scan of the SNR for different values of the amplitude $A$ and $f_*$ for $c_s^2=w$ fixing $f_{\rm RH} = 10^{-5}\,\mathrm{Hz}$. To illustrate the effect of different reheating scales, we also show the lines corresponding to $f_{\rm RH} = 10^{-7}\,\mathrm{Hz}$ (dashed) and $f_{\rm RH} = 10^{-9}\,\mathrm{Hz}$ (dotted) for $\mathrm{SNR} = 1$ in the first case and for $\mathrm{SNR} = 10$ in the other two cases. The grey shaded region shows the BBN bounds on GWs \cite{Caprini:2018mtu}. \textbf{The top Left Panel} shows the case $w = 0.1$ and the solid lines correspond to $\mathrm{SNR} = 1$ and $100$ for reference.  
    \textbf{The top Right Panel} shows the case $w = 0.5$ and the solid lines correspond to $\mathrm{SNR} = 10$ and $10^4$ for reference.  
    \textbf{The bottom Panel} shows $w = 0.9$ and the solid lines correspond to $\mathrm{SNR} = 10$ and $10^4$ for reference.
}
    \label{fig:scans}
\end{figure}

The results of the previous section demonstrated LISA’s capability to detect a SIGW signal generated during the reheating phase for different benchmark values of $w$ and reheating frequencies (or, equivalently, temperatures). To explore the detection capability across the full parameter space, we evaluate the SNR for different combinations of the signal amplitude and peak frequency. In all three cases, we fix the reheating frequency to $f_{\rm RH}=10^{-5}\,\mathrm{Hz}$ for illustrative purposes, while also displaying lines corresponding to different values of $f_{\rm RH}$ to provide intuition on how the contours shift. Our results are shown in \cref{fig:scans}. To account for the BBN bounds, we shade all regions where the GW signal would exceed the value $\sim 10^{-6}$.\footnote{The shaded area shifts as $f_{\rm RH}$ varies. A qualitative estimate of the corresponding boundary can be obtained from
\be
A = \sqrt{\frac{h^2\Omega_{\rm thr}}{h^2\Omega_{\rm rad,0} c_g , \Omega_{\rm max}}} \left(\frac{f_*}{f_{\rm RH}}\right)^b \,,
\ee
with $h^2\Omega_{\rm thr}\sim10^{-6}$~\cite{Caprini:2018mtu}, $c_g = 0.39$~\cite{Borsanyi:2016ksw} and $h^2\Omega_{\rm rad,0}\simeq4.2\cdot10^{-5}$~\cite{Planck:2018vyg}. $\Omega_{\rm max}$ is instead the value of the spectrum in the peak, which is usually of order 1 (see left panel of~\cref{fig:Gaussian_Comparison}). Note that for sufficiently narrow spectra, this approximation is reliable, since all of the power is mainly concentrated around the peak.}
For different combinations of amplitudes and peak frequencies, we evaluate the corresponding SNR. The SNR is defined as
\bea
\label{Eq:SNR}
    {\rm SNR} = \sqrt{T_{\rm obs}} \sqrt{\int df \left( \frac{h^2 \Omega_{\rm GW}(f)}{h^2 \Omega_{\rm noise}(f)}\right)^2}\,,
\eea
where
\bea
    h^2 \Omega_{\rm noise}(f) = \frac{4 \pi^2 f^3}{3 H_0^2/h^2}S_h(f)\,,
\eea
and $S_h(f)$ is the LISA sensitivity \cite{Babak:2021mhe,LISACosmologyWorkingGroup:2024hsc,LISACosmologyWorkingGroup:2025vdz}. We also include foregrounds due to extragalactic and galactic binaries, adding them as a further source of noise, following~\cite{LISACosmologyWorkingGroup:2024hsc,LISACosmologyWorkingGroup:2025vdz}. In each panel, we report contour lines corresponding to different SNR values for reference. We again discuss the cases $w=0.1$ and $w=\{0.5,0.9\}$ separately.

First, see the top-left plot of \cref{fig:scans} which corresponds to $w=0.1$. In that case, we choose as references the contours corresponding to $\mathrm{SNR}=1$ and $\mathrm{SNR}=10^2$. Because of the suppression for $b>0$ (see eq.~\eqref{eq:frh_dep}), the SNR values are typically lower than in the other cases for a given $(A,f_*)$.
For $\mathrm{SNR}=1$, the solid line corresponds to $f_{\rm RH} = 10^{-5}$ Hz, while the dashed and dotted lines correspond to $f_{\rm RH} = 10^{-7}$ Hz and $f_{\rm RH} = 10^{-9}$ Hz, respectively. These lines illustrate the shift in the SNR as the reheating frequency varies. In particular, as expected, when $f_{\rm RH}$ increases, the curves at fixed SNR shift upwards. In other words, a larger amplitude is required to achieve the same SNR due to the suppression caused by the evolution of the Universe. The impact of foregrounds can be seen in the minima of the curves, which are responsible for the bump around milliHz frequencies.

For $w=0.5$ (top-right panel) and $0.9$ (bottom panel), the results are opposite. Due to the spectrum enhancement, all frequencies are boosted, so we plot the contours corresponding to $\mathrm{SNR}=10$ and $\mathrm{SNR}=100$. Notably, a detection of SIGWs with $\mathrm{SNR}=10$ would be possible even with a very low primordial spectrum amplitude, of the order of $A\sim 10^{-5}$. As $f_{\rm RH}$ decreases, the curves at fixed SNR shift downward, indicating that even spectra with lower amplitudes could be observed due to reheating-induced boosting. As $w$ varies, we observe that both the left and right parts of the contour lines change slope. This is due to the variation of specific IR and UV features with $w$, as discussed in \cref{sec:spectra_eval}. One can also notice that the shaded region shifts downward with increasing $w$, because the larger $w$ is compared to $1/3$, the stronger the boost, allowing lower amplitudes to more easily saturate the BBN bounds. 

An interesting implication concerns the abundance of primordial black holes (PBHs) produced. Some of the primordial perturbations responsible for generating SIGWs, if sufficiently large in amplitude, can lead to PBH formation~\cite{Zeldovich:1967lct,Hawking:1971ei,Carr:1974nx,Carr:1975qj,Chapline:1975ojl,Carr:2026hot}. Typically, the PBH abundance is related to the primordial spectrum amplitude, the level of non-Gaussianity, and the peak frequency (see e.g.~\cite{Young:2022phe,Ferrante:2022mui}). However, as we have shown, in the case of reheating, the value of $w$ is crucial. When the spectrum is boosted, much lower primordial amplitudes are sufficient to produce an observable SIGW spectrum. Lower amplitudes naively correspond to a smaller PBH abundance, meaning that gravitational waves could be potentially observed without necessarily producing an excessive number of PBHs. Conversely, when gravitational waves are suppressed, the opposite reasoning applies. See, e.g., \cite{Domenech:2020ers,Zhu:2023gmx,Liu:2023pau,Harigaya:2023pmw,Domenech:2024rks} for analyses considering a general equation of state in the Gaussian case in the PTA band. Since a detailed analysis of PBH abundance depends on the specific model considered, in particular the value of $w$ and $f_{\rm NL}$, we leave such a study for future work.

\section{Conclusions}
\label{sec:conclusion}
We studied for the first time the impact of primordial NG on the SIGW spectrum during non-standard reheating epochs. We considered a few physically motivated models, like cannibal dark matter ($w=0.1$) \cite{Erickcek:2020wzd}, chaotic models ($w\sim0.2$) \cite{Podolsky:2005bw}, power-law oscillations with fragmentation ($w=0.5$) \cite{Bernal:2019mhf,Garcia:2020eof, Garcia:2023dyf}, and almost stiff matter ($w=0.9$) \cite{Spokoiny:1993kt}. In \cref{fig:spectra,fig:Gaussian_Comparison}, we show the SIGW spectra for different values of $w$ in both the Gaussian and non-Gaussian cases. We argue that the effects of the equation of state can be inferred even without observing the full spectrum, due to the imprints of $w$. The specific features can already be seen in the Gaussian case, depending on the chosen values of $w$ and $c_s^2$, as shown in both the panels of \cref{fig:Gaussian_Comparison}. These results suggest the possibility of discriminating among different models by observing the SIGW signal produced. In addition, although we assumed the standard polynomial expansion for primordial non-Gaussianity for simplicity, our results can be generalised to any non-Gaussian shape, provided an appropriate trispectrum is used. The kernel in \cref{eq:generalized_kernel} remains unchanged.

Our findings are further supported by Fisher forecasts that we perform for different values of $w$. Depending on the values of $w$ and on the reheating frequency, the spectrum can be either enhanced or suppressed. Specifically, for $w<1/3$, the spectrum is suppressed even for large values of $A$, making the parameter reconstruction more challenging. Nevertheless, the relative errors on $w$ and $c_s^2$ remain the lowest, reflecting the distinct spectral features they imprint. On the contrary, when $w>1/3$, the spectrum is enhanced and detectable even for lower amplitudes, making parameter reconstruction more feasible. This is confirmed by scans in the parameter space over different amplitudes and central frequencies for different values of $w$. When the spectrum is enhanced, high SNR values can be achieved even for amplitudes as low as $10^{-5}$.
Our results show that, within the specific model and assumptions adopted in this work, a detection of a SIGW signal can provide precise constraints on $w$, $c_s$, and $f_{\rm NL}$. The extent to which these constraints can be obtained in more general scenarios depends on the underlying model and assumptions. Most importantly, leaving an arbitrary value of $w$ in the analysis does not hinder the detection of primordial NG, as they are responsible for different features in the SIGW spectrum. Furthermore, our formalism lays the ground for a very general analysis of current and future GW data regarding SIGWs.

An interesting follow-up analysis concerns the production of primordial black holes in these regimes, considering different values of $w$ and $f_{\rm NL}$. When the SIGW signal is enhanced, GWs can be detected even for a low primordial amplitude. However, this simultaneously suppresses the production of PBHs, possibly limiting their observability.

\section*{Acknowledgements}
G.P. thanks Fondazione Angelo Della Riccia and Fondazione Aldo Gini and the Estonian Research Council grants KOHTO34, TARISTU24-TK10, TARISTU24-TK3, and the CoE TK 202 ``Fundamental Universe’' for financial support. G.P. thanks the Institute for Theoretical Physics at Leibniz University Hannover for the hospitality. G.D. is supported by the DFG under the Emmy-Noether program, project number 496592360, and by the JSPS KAKENHI grant No. JP24K00624.

%%%%%%%%%%%%%%%%%%%%%%%%%%
\appendix
\section{Evolution of scalar perturbations}
\label{app:scalars_eom}
Scalar perturbations at first order in perturbation theory constitute the source for GWs at second order. We need to get the evolution of first-order scalar perturbations. Starting from the Einstein-Equations using Eq. \eqref{eq:metric}, in the absence of anisotropic stress, one obtains
\be
\Phi''+3\mathcal{H}(1+c_w^2)\Phi'+(2\mathcal{H}'+(1+3 c_w^2)\mathcal{H}^2)\Phi+c_s^2k^2\Phi = 0\,.
\ee
Here the prime indicates a derivative with respect to the conformal time $\eta$, and we defined $c_w^2 = \dot{P}/\dot{\rho}$ and $c_s^2 = \delta {P}/\delta {\rho}$. For an adiabatic perfect fluid, these two quantities are equal and correspond to $w$. The equation can be solved analytically when $w$ and $c_s^2$ are constant: by imposing that the solution must be constant on superhorizon scales, one is able to write~\cite{Mukhanov:2005sc,Domenech:2021ztg}
\be
\Phi(k\eta)=\Phi(\vec{k}) 2^{b+\frac{3}{2}}\Gamma\left[b+\frac{5}{2}\right](c_sk\eta)^{-b-\frac{3}{2}}J_{b+\frac{3}{2}}(c_s k\eta)\,,
\ee
with $J_{\nu}$ Bessel function of the first kind and $\Phi(\vec{k})$ the value of the field at the initial time. The last equation allows us to define the transfer function $T_\Phi(k\eta)$ as
\bea
\label{eq:tphi}
    T_\Phi(k\eta) = & 2^{b+\frac{3}{2}}\Gamma\left[b+\frac{5}{2}\right](c_sk\eta)^{-b-\frac{3}{2}}J_{b+\frac{3}{2}}(c_s k\eta)\\
    = & 2^{b+\frac{3}{2}} \Gamma\left[b+\frac{5}{2}\right] \frac{(c_s k \eta)^{-b-\frac{1}{2}}}{3+2b}\left(J_{b+\frac{1}{2}}(c_s k \eta)+J_{b+\frac{5}{2}}(c_s k \eta)\right)\,,
\eea
where in the last line we used the recursive relation for Bessel functions. Additionally
\bea
\label{eq:tphiprime}
    T_\Phi'(k,\eta) = \partial_\eta T_\Phi(k,\eta) = & -{2^{b+\frac 3 2}}\Gamma\left[b+\frac 5 2 \right] \frac{(c_s k \eta)^{-b-\frac{1}{2}}}{\eta} J_{b+\frac{5}{2}}(c_s k \eta)\,.
\eea

\section{Green's method for tensors}
\label{app:Greens_tensors}
In the main text we need to solve the equation of motion for the tensor perturbation $h$, when a source term is present. For an equation
\be
h_{k,\lambda}''(k,\eta)+2\mathcal{H}h'(k,\eta)+k^2h(k,\eta)={S}_{\lambda}(\bar{\eta},k)\,,
\ee
the solution can be written using the Green's method as
\be
h_{\lambda}(k,\eta)=\int_0^\eta d\bar{\eta}G_{k,\lambda}(\eta,\bar{\eta}) S_\lambda(\bar{\eta},k)\,,
\ee
with initial conditions $h_\lambda(k,\eta_{\rm i})=h'_\lambda(k,\eta_{\rm i})=0$. In the last equation $G_{k,\lambda}(\eta,\bar{\eta})$ is the Green's function and solves 
\be
G_{k,\lambda}(\eta,\bar{\eta})''+2\mathcal{H}G'_{k,\lambda}(\eta,\bar{\eta})+k^2G_{k,\lambda}(\eta,\bar{\eta})=\delta(\eta-\bar{\eta})\,.
\ee
It can be obtained starting from the two homogeneous solutions, which we call $h_1(k,{\eta})$ and $h_2(k,{\eta})$, as
\be
G_{k,\lambda}(\eta,\bar{\eta}) = \frac{1}{W[h_1(k,\bar{\eta}),h_2(k,\bar{\eta})]}[h_1(k,{\eta}),h_2(k,\bar{\eta})-h_2(k,{\eta}),h_1(k,\bar{\eta})]\,,
\ee
with $W$ the Wronskian defined as 
\be
W[h_1(k,\bar{\eta}),h_2(k,\bar{\eta})] = h'_1(k,\bar{\eta})h_2(k,\bar{\eta})-h'_2(k,\bar{\eta})h_1(k,\bar{\eta})\,.
\ee
For a generic equation of state $w$ one obtains
\be
h_1(k,\eta)=(k\eta)^{-b-\frac{1}{2}} J_{b+\frac{1}{2}}(k\eta) \quad h_2(k,\eta)=(k\eta)^{-b-\frac{1}{2}} Y_{b+\frac{1}{2}}(k\eta)\,,
\ee
where again $J_\nu$ and $Y_\nu$ are the Bessel function of the first and second kind. The Green's function thus results
\be
\label{eq:green_k}
    G_{k,\lambda}(\eta,\bar{\eta}) = \frac{\pi}{2k} \frac{(k\bar{\eta})^{b+\frac{3}{2}}}{(k{\eta})^{b+\frac{1}{2}}} \left(J_{b+\frac{1}{2}}(k\bar{\eta})Y_{b+\frac{1}{2}}(k{\eta})-Y_{b+\frac{1}{2}}(k\bar{\eta})J_{b+\frac{1}{2}}(k{\eta})\right)\,.
\ee
\section{Tensor power spectra}
\label{app:Psh}
As argued in the main text, the tensor power spectrum for SIGWs depends on the trispecturm of scalar perturbations. Starting from~\cref{eq:tensor_h_sigw_2,eq:tensor_ps} and substituting~\eqref{eq:local_exp}, after performing all the Wick contractions one obtains for the Gaussian case
\begin{equation}
    \begin{aligned}
    P_{h,\lambda}(k,\eta)\big|_{\rm G} = 2^5 \int \frac{d^3 q}{(2\pi)^3}Q^2_{\lambda}(\vec{k},\vec{q})\mathbb{I}^2(|\vec{k}-\vec{q}|,q,\eta) P_{\zeta_{\rm G}}(q)P_{\zeta_{\rm G}}(|\vec{k}-\vec{q}|)\,;
    \end{aligned}
    \label{eq:Gaussian}
\end{equation}
The $f_{\rm NL}$ contributions, instead, appear proportional only to the square of this parameter. This is related to the fact that only an even number of Gaussian fields with vanishing mean, if contracted, lead to a non-vanishing contribution to the trispectrum. Thus, we obtain for the connected part 
\begin{equation}
    \begin{aligned}
    P_{h,\lambda}(k,\eta)\big|_{\rm t} =& 2^8 \left(\frac{3}{5}f_{\rm NL}\right)^2 \int \frac{d^3 q_1}{(2\pi)^3} \int \frac{d^3 q_2}{(2\pi)^3}Q_{\lambda}(\vec{k},\vec{q}_1)Q_{\lambda}(\vec{k},\vec{q}_2)\\
    &\times\mathbb{I}(|\vec{k}-\vec{q}_1|,q_1,\eta)\mathbb{I}(|\vec{k}-\vec{q}_2|,q_2,\eta)  P_{\zeta_{\rm G}}(q_2)P_{\zeta_{\rm G}}(|\vec{k}-\vec{q}_2|)P_{\zeta_{\rm G}}(|\vec{q}_1-\vec{q}_2|)\,,
    \end{aligned}
    \label{t}
\end{equation}

\begin{equation}
    \begin{aligned}
    P_{h,\lambda}(k,\eta)\big|_{\rm u} =& \hspace{0.1cm} 2^8 \left(\frac{3}{5}f_{\rm NL}\right)^2 \int \frac{d^3 q_1}{(2\pi)^3} \int \frac{d^3 q_2}{(2\pi)^3}Q_{\lambda}(\vec{k},\vec{q}_1)Q_{\lambda}(\vec{k},\vec{q}_2) \\
    &\times\mathbb{I}(|\vec{k}-\vec{q}_1|,q_1,\eta)\mathbb{I}(|\vec{k}-\vec{q}_2|,q_2,\eta) P_{\zeta_{\rm G}}(q_1)P_{\zeta_{\rm G}}(q_2)P_{\zeta_{\rm G}}(|\vec{k}-(\vec{q}_1+\vec{q}_2)|)\,,
    \end{aligned}
    \label{u}
\end{equation}
\begin{equation}
    \begin{aligned}
    P_{h,\lambda}(k,\eta)\big|_{\rm s} =& \hspace{0.1cm} 2^8 \left(\frac{3}{5}f_{\rm NL}\right)^2 \int \frac{d^3 q_1}{(2\pi)^3} \int \frac{d^3 q_2}{(2\pi)^3}Q_{\lambda}(\vec{k},\vec{q}_1)Q_{\lambda}(\vec{k},\vec{q}_2)\\
    &\times\mathbb{I}(|\vec{k}-\vec{q}_1|,q_1,\eta)\mathbb{I}(|\vec{k}-\vec{q}_2|,q_2,\eta) P_{\zeta_{\rm G}}(q_1)P_{\zeta_{\rm G}}(q_2)P_{\zeta_{\rm G}}(k)\,,
    \end{aligned}
    \label{s}
\end{equation}
and for the disconnected part
\begin{equation}
    \begin{aligned}
    P_{h,\lambda}(k,\eta)\big|_{\rm hyb} = & \hspace{0.1cm} 2^7 \left(\frac{3}{5}f_{\rm NL}\right)^2 \int \frac{d^3 q_1}{(2\pi)^3}\int \frac{d^3 q_2}{(2\pi)^3}Q^2_{\lambda}(\vec{k},\vec{q}_1)\\
    &\times\mathbb{I}^2(|\vec{k}-\vec{q}_1|,q_1,\eta) P_{\zeta_{\rm G}}(|\vec{k}-\vec{q}_1|)P_{\zeta_{\rm G}}(q_2)P_{\zeta_{\rm G}}(|\vec{q}_1-\vec{q}_2|)\,.
    \end{aligned}
    \label{ibrid}
\end{equation}
We start from these expressions to obtain the final expression for $\Omega_{\rm GW}(k)$ in the various cases. As a further remark, we note that although all the trispectra we report do not vanish, the final tensor power spectrum can still vanish. This is the case for the s contribution, as shown in~\cite{Unal:2018yaa,Adshead:2021hnm,Perna:2024ehx}. In addition, there are two additional contributions at the next-to-leading order in perturbation theory proportional to $g_{\rm NL}$ (i.e. the parameter associated with $\zeta_{\rm G}^3$ in the expansion). However, as shown in~\cite{Perna:2024ehx}, one of those two contributions vanishes, while the other is fully degenerate with the Gaussian and thus does not add any additional feature to the GW spectrum. For the purposes of this analysis, it is thus safe to put $g_{\rm NL}=0$ and neglect those terms. 
\section{Comparison of the Kernel}
\label{app:comparison}
In this appendix, we evaluate the kernel in Eq.~\eqref{eq:general_kernel} in the usual case with $c_s^2=w=1/3$ to compare our result with previous works. Since $y= \frac{u^2+v^2-3}{2uv}$, we get
\bea
\left(\mathsf{P}_b^{-b}(y)+\frac{b+2}{b+1}\mathsf{P}_{b+2}^{-b}(y)\right)\bigg|_{b=0} = &\left(1+2 \frac{3y^2-1}{2}\right)= 3 y^2\,,\\
\left(\mathsf{Q}_b^{-b}(y)+\frac{b+2}{b+1}\mathsf{Q}_{b+2}^{-b}(y)\right)\bigg|_{b=0} = & 3y \left[\frac{y}{2}\ln\left(\frac{1+y}{1-y}\right)-1\right]\,,\\
\left(\mathcal{Q}_{b}^{-b}(-y)+2\frac{b+2}{b+1}\mathcal{Q}_{b+2}^{-b}(-y)\right)\bigg|_{b=0} = & -3y\left[\frac{y}{2}\ln\left(\frac{y+1}{y-1}\right)-1\right]\,,
\eea
and thus
\bea
I(v,u,x\gg1) & = \frac{2}{k^2}\frac{3}{2} \frac{\pi}{4}x^{-1}\frac{3}{u v}3y \left\{-y \cos x \Theta(u+v-\sqrt{3}) + \sin x \frac{2}{\pi}\left[\frac{y}{2}\ln\left |\frac{y+1}{y-1}\right|-1\right]\right\}\\
& = \frac{2}{k^2}\frac{(3)}{(2)} \frac{1}{4}x^{-1}\frac{3}{u v}3 \frac{u^2+v^2-3}{(2uv)^2}\Bigg\{-\cos x \pi (u^2+v^2-3)  \Theta(u+v-\sqrt{3})\\
&
  \qquad\qquad+ \sin x \left[(u^2+v^2-3)\ln\left |\frac{(u+v)^2-3}{(u-v)^2-3}\right|-4uv\right]\Bigg\}\,.
\eea
This expression coincides with eq.~(25) of \cite{Kohri:2018awv}, taking into account that they use $\mathbb{I}$ and, according to their convention, they define the kernel without the $1/k^2$ factor. In fact, squaring and performing the oscillation average, we obtain
\bea
\overline{{\mathbb{I}(v,u,x\gg1)}^2} = &\frac{1}{2 k^4x^2}\left(\frac{3(u^2+v^2-3)}{4(uv)^3}\right)^2\Bigg\{ \pi^2 (u^2+v^2-3)^2  \Theta(u+v-\sqrt{3}) \\
&\qquad\qquad+ \left[(u^2+v^2-3)\ln\left |\frac{(u+v)^2-3}{(u-v)^2-3}\right|-4uv\right]^2\Bigg\}\,,
\eea
which, apart from the factor $1/k^4$ coincides with their eq.~(26).

\section{Useful functions}
\label{app:Olv_Ferr}
Ferrer's and Olver's functions are special functions which, for integer values of $b$, can be written analytically. This helps avoid numerical divergences or artefacts that could arise. In our case we have some particular values of $w$ which lead to integer (or half integer) $b$, i.e. $w=1/3\to b=0$, $w=0\to b=1$, $w=1\to b=-1/2$. We thus report first the general definitions of those functions and then some special cases (see also~\cite{Domenech:2020kqm}).
We have
\bea
\mathsf{P}^{\mu}_{\nu}\left(x\right)=\left(\frac{1+x}{1-x}\right)^{\mu/2}%
\mathbf{F}\left(\nu+1,-\nu;1-\mu;\tfrac{1}{2}-\tfrac{1}{2}x\right)\,,
\eea
\bea
\mathsf{Q}^{\mu}_{\nu}\left(x\right)=&\frac{\pi}{2\sin\left(\mu\pi\right)}\Bigg%
(\cos\left(\mu\pi\right)\left(\frac{1+x}{1-x}\right)^{\mu/2}\mathbf{F}\left(%
\nu+1,-\nu;1-\mu;\tfrac{1}{2}-\tfrac{1}{2}x\right)\nonumber\\&\qquad\qquad\qquad-\frac{\Gamma\left(\nu+\mu+1%
\right)}{\Gamma\left(\nu-\mu+1\right)}\left(\frac{1-x}{1+x}\right)^{\mu/2}%
\mathbf{F}\left(\nu+1,-\nu;1+\mu;\tfrac{1}{2}-\tfrac{1}{2}x\right)\Bigg)\,,
\eea
and
\bea
{\cal Q}^{\mu}_{\nu}\left(x\right)=&%
\frac{\pi}{2\sin\left(\mu\pi\right)\Gamma\left(\nu+\mu+1\right)}\Bigg(\left(\frac{x+1}{x-1}\right)^{\mu/2}\mathbf{F}\left%
(\nu+1,-\nu;1-\mu;\tfrac{1}{2}-\tfrac{1}{2}x\right)\nonumber\\&\qquad\qquad\qquad-\frac{\Gamma\left(\nu+\mu+1\right)}{%
\Gamma\left(\nu-\mu+1\right)}\left(\frac{x-1}{x+1}\right)^{\mu/2}\mathbf{F}\left(\nu+1,-\nu;1+\mu;%
\tfrac{1}{2}-\tfrac{1}{2}x\right)\Bigg)\,.
\eea
Note that the Ferrer's functions are valid for $|x|<1$, while the Olver's function is real for $|x|>1$. In the above definitions, we used the compact notation
\bea
\mathbf{F}\left(a,b;c;x\right)=\frac{1}{\Gamma[c]}{}_2{F}_1\left(a,b;c;x\right)\,,
\eea
with ${}_2{F}_1\left(a,b;c;x\right)$ being the Gauss's Hypergeometric function. In our case, we need to evaluate these functions for $\nu=-\mu=b$ and $\mu=-b$, $\nu=b+2$.
\paragraph{$\bullet\quad\bf b=0$}
\bea
    &\mathsf{P}_0^0(x) = 1\,, & & \mathsf{P}^0_2 = \frac{3x^2-1}{2} \,, \\
    &\mathsf{Q}_0^0(x) = \frac{1}{2} \ln \left(\frac{1+x}{1-x}\right)\,,  & & \mathsf{Q}_2^0(x) = \frac{3x^2 - 1}{4} \ln \left( \frac{1+x}{1-x}\right) - \frac{3}{2} x\,, \\
    & \mathcal{Q}_0^0(x) = \frac{1}{2} \ln \left(\frac{x+1}{x-1}\right)\,, & & \mathcal{Q}_2^0(x) = \frac{3x^2 - 1}{8} \ln \left( \frac{x+1}{x-1}\right) - \frac{3}{4} x\,. \\
\eea
\paragraph{$\bullet\quad \bf b=1$}
\bea
    &\mathsf{P}_{1}^{-1} (x) = \frac{\sqrt{1-x^2}}{2}\,, & & \mathsf{P}^{-1}_3 = \left(\frac{1+x}{1-x}\right)^{-\frac{1}{2}}
    \frac{1}{8} (1+x)(5x^2-1)\,, \\
    &\mathsf{Q}_{1}^{-1}(x) = -\frac{\left(x^2-1\right) \ln \left(\frac{x+1}{1-x}\right)-2 x}{4 \sqrt{1-x^2}}\,,  & &  \mathsf{Q}_{3}^{-1}(x) = -\frac{-30 x^3+3 \left(5 x^4-6 x^2+1\right) \log \left(\frac{x+1}{1-x}\right)+26 x}{48 \sqrt{1-x^2}}\,, \\
    & \mathcal{Q}_1^{-1}(x) = \frac{2 x-\left(x^2-1\right) \ln \left(\frac{x+1}{x-1}\right)}{4 \sqrt{x^2-1}}\,, & & \mathcal{Q}_{2}^{0}(x) = \frac{\left(3 x^2-1\right)}{8} \ln \left(\frac{x+1}{x-1}\right)-\frac{3}{4} x\,.
\eea
\paragraph{$\bullet\quad\bf b=-1/2$}
\bea
    &\mathsf{P}^{1/2}_{-1/2}(x) = \left(\frac{2}{\pi\sqrt{(1-x^2)}}\right)^{\frac{1}{2}}\,, & & \mathsf{P}^{1/2}_{3/2}(x) =\left(\frac{2}{\pi\sqrt{(1-x^2)}}\right)^{\frac{1}{2}} (2x^2-1)\,,\\
    &\mathsf{Q}^{1/2}_{-1/2}\left(x\right)=0\,, & & \mathsf{Q}^{1/2}_{3/2}\left(x\right) = -\sqrt{2 \pi } x \sqrt[4]{1-x^2}\,, \\
    & \mathcal{Q}^{+ 1/2}_{-1/2}\left(x\right)= \left(\frac{\pi}{2(x^2-1)^{1/2}}\right)^{1/2}\,,& & {\mathcal{Q}^{\pm 1/2}_{3/2}\left(x\right)} = \left(\frac{\pi}{2(x^2-1)^{1/2}}\right)^{1/2} \frac{1}{2\left(x+\sqrt{x^2-1}\right)^2}\,.
\eea
Further properties can be found in~\cite{DLMF14.3}.
\bibliographystyle{JHEP}
\bibliography{refs}

\end{document}